\documentclass[iop,revtex4.1,numberedappendix,appendixfloats]{emulateapj}

\usepackage{natbib}
\usepackage{amsmath}
\usepackage{color}
\usepackage{url}
\usepackage{ulem}
\usepackage{multirow}
\usepackage{amsmath}
\usepackage{wasysym}

\bibpunct{(}{)}{;}{a}{}{,}

\def\apjl{ApJ}%
\def\apjs{ApJS}%
\def\ao{Appl.~Opt.}%
\def\aap{A\&A}%
\newcommand{\sname}{WISE\,J1036+0449}

\newcommand\C   {\textsc{Clumpy}}        
\newcommand\mic {\hbox{$\mu{\rm m}$}}

\shorttitle{ {\it NuSTAR} observations of a Hot, Dust Obscured Galaxy at $z\sim1$}
\shortauthors{Ricci et al.}

\begin{document}

\title{{\it NuSTAR} observations of WISE\,J1036+0449, a Galaxy at \small{z}\normalsize{$\sim1$ obscured by hot dust}

}

\author{C. Ricci\altaffilmark{1,2,3,*}, R. J. Assef\altaffilmark{4}, D. Stern\altaffilmark{5}, R. Nikutta\altaffilmark{1,6}, D. M. Alexander\altaffilmark{7}, D. Asmus\altaffilmark{8},  D. R. Ballantyne\altaffilmark{9}, F. E. Bauer\altaffilmark{1,2,10,11},  A. W. Blain\altaffilmark{12}, S. Boggs\altaffilmark{13}, P. G. Boorman\altaffilmark{14}, W. N. Brandt\altaffilmark{15,16,17}, M. Brightman\altaffilmark{18}, C. S. Chang\altaffilmark{19}, C.-T. J. Chen\altaffilmark{15}, F. E. Christensen\altaffilmark{20},  A. Comastri\altaffilmark{21}, W. W. Craig\altaffilmark{13}, T. D\'iaz-Santos\altaffilmark{4}, P. R. Eisenhardt\altaffilmark{5},  D. Farrah\altaffilmark{22}, P. Gandhi\altaffilmark{14},  C. J. Hailey\altaffilmark{23}, F. A. Harrison\altaffilmark{18}, H. D. Jun\altaffilmark{5}, M. J. Koss\altaffilmark{24,25}, S. LaMassa\altaffilmark{26},  G. B. Lansbury\altaffilmark{7}, C. B. Markwardt\altaffilmark{27,28}, M. Stalevski\altaffilmark{29,30,31}, F. Stanley\altaffilmark{7}, E. Treister\altaffilmark{1,2},  C.-W.Tsai\altaffilmark{5}, D. J. Walton\altaffilmark{5,32}, J. W. Wu\altaffilmark{33}, L. Zappacosta\altaffilmark{34}, W. W. Zhang\altaffilmark{28}
}

\altaffiltext{1}{Instituto de Astrof\'{\i}sica, Facultad de F\'{i}sica, Pontificia Universidad Cat\'{o}lica de Chile, Casilla 306, Santiago 22, Chile} 
\altaffiltext{2}{EMBIGGEN Anillo, Concepcion, Chile}
\altaffiltext{3}{Kavli Institute for Astronomy and Astrophysics, Peking University, Beijing 100871, China}
\altaffiltext{4}{N\'ucleo de Astronom\'ia de la Facultad de Ingenier\'ia, Universidad Diego Portales, Av. Ej\'ercito Libertador 441, Santiago, Chile}
\altaffiltext{5}{Jet Propulsion Laboratory, California Institute of Technology, Pasadena, CA 91109, USA}
\altaffiltext{6}{National Optical Astronomy Observatory, 950 N Cherry Ave, Tucson, AZ 85719, USA}
\altaffiltext{7}{Centre for Extragalactic Astronomy, Department of Physics, Durham University, South Road, Durham, DH1 3LE, UK}
\altaffiltext{8}{European Southern Observatory, Casilla 19001, Santiago 19, Chile} 
\altaffiltext{9}{Center for Relativistic Astrophysics, School of Physics, Georgia Institute of Technology, 837 State Street, Atlanta, GA 30332-0430, USA}
\altaffiltext{10}{Space Science Institute, 4750 Walnut Street, Suite 205, Boulder, Colorado 80301, USA}
\altaffiltext{11}{Millenium Institute of Astrophysics, Santiago, Chile}
\altaffiltext{12}{University of Leicester, Physics \& Astronomy, 1 University Road, Leicester LE1 7RH, UK}
\altaffiltext{13}{ Space Sciences Laboratory, University of California, Berkeley, CA 94720, USA}
\altaffiltext{14}{Department of Physics and Astronomy, University of Southampton, Highfield, Southampton SO17 1BJ, UK}
\altaffiltext{15}{Department of Astronomy and Astrophysics, The Pennsylvania State University, University Park, PA 16802, USA}
\altaffiltext{16}{Institute for Gravitation and the Cosmos, The Pennsylvania State University, University Park, PA 16802, USA}
\altaffiltext{17}{Department of Physics, 104 Davey Lab, The Pennsylvania State University, University Park, PA 16802, USA}
\altaffiltext{18}{Cahill Center for Astronomy and Astrophysics, California Institute of Technology, Pasadena, CA 91125, USA }
\altaffiltext{19}{Joint ALMA Observatory, Alonso de Cordova 3107, Vitacura, Santiago, Chile}
\altaffiltext{20}{DTU Space, National Space Institute, Technical University of Denmark, Elektronvej 327, DK-2800 Lyngby, Denmark}
\altaffiltext{21}{INAF -- Osservatorio Astronomico di Bologna, via Ranzani 1, 40127 Bologna, Italy}
\altaffiltext{22}{Department of Physics, Virginia Tech, Blacksburg, VA 24061, USA}
\altaffiltext{23}{Columbia Astrophysics Laboratory, Columbia University, New York 10027, USA}
\altaffiltext{24}{Institute for Astronomy, Department of Physics, ETH Zurich, Wolfgang-Pauli-Strasse 27, CH-8093 Zurich, Switzerland}
\altaffiltext{25}{Ambizione fellow}
\altaffiltext{26}{NASA Goddard Space Flight Center, Greenbelt, MD 20771, USA}
\altaffiltext{27}{Department of Astronomy, University of Maryland, College Park, MD 20742, USA}
\altaffiltext{28}{Astroparticle Physics Laboratory, Mail Code 661, NASA Goddard Space Flight Center, Greenbelt, MD 20771, USA}
\altaffiltext{29}{Departamento de Astronomia, Universidad de Chile, Camino El Observatorio 1515, Casilla 36-D Santiago, Chile}
\altaffiltext{30}{Astronomical Observatory, Volgina 7, 11060 Belgrade, Serbia}
\altaffiltext{31}{Sterrenkundig Observatorium, Universiteit Gent, Krijgslaan 281-S9, Gent B-9000, Belgium}
\altaffiltext{32}{Space Radiation Laboratory, California Institute of Technology, Pasadena, CA 91125, USA}
\altaffiltext{33}{National Astronomical Observatories, Chinese Academy of Sciences, 20A Datun Road, Chaoyang District, Beijing, 100012, China}
\altaffiltext{34}{INAF -- Osservatorio Astronomico di Roma, via Frascati 33, 00078 Monte Porzio Catone (RM), Italy}

\altaffiltext{*}{cricci@astro.puc.cl}

\begin{abstract}
Hot, Dust-Obscured Galaxies (Hot DOGs),  selected from the {\it WISE} all sky infrared survey, host some of the most powerful
Active Galactic Nuclei (AGN) known, and might represent an important
stage in the evolution of galaxies. Most known Hot DOGs are at $z> 1.5$, due in part to a strong
bias against identifying them at lower redshift related to the selection criteria. We present a new selection method that identifies 153 Hot DOG candidates at $z\sim 1$, where they are
significantly brighter and easier to study. We validate this approach by measuring a redshift $z=1.009$, and an SED similar to higher redshift Hot DOGs for one of these objects, WISE\,J1036+0449 ($L_{\rm\,Bol}\simeq 8\times 10^{46}\rm\,erg\,s^{-1}$), using data from Keck/LRIS and NIRSPEC, SDSS, and CSO. We find evidence of a broadened component in Mg\,{\sc ii}, which, if due to the gravitational potential of the supermassive black hole, would imply a black hole mass of $M_{\rm\,BH}\simeq 2 \times 10^8 M_{\odot}$, and an Eddington ratio of $\lambda_{\rm\,Edd}\simeq 2.7$. WISE\,J1036+0449 is the first Hot DOG detected by {\it NuSTAR}, and the observations show that the source is heavily
obscured, with a column density of $N_{\rm\,H}\simeq(2-15)\times10^{23}\rm\,cm^{-2}$. The source has an intrinsic
2--10\,keV luminosity of $\sim 6\times 10^{44}\rm\,erg\,s^{-1}$, a value significantly lower than that expected from
the mid-infrared/X-ray correlation. We also find that the other Hot DOGs observed by X-ray facilities show a similar deficiency of X-ray flux. We discuss the origin of the X-ray weakness and the absorption properties of Hot DOGs.
Hot DOGs at $z\lesssim1$ could be
excellent laboratories to probe the characteristics of the accretion
flow and of the X-ray emitting plasma at extreme values of the Eddington
ratio.
\end{abstract}


\section{Introduction}

\setcounter{footnote}{0}

Supermassive black holes (SMBHs) are known to reside at the centers of
most galaxies (e.g., \citealp{Kormendy:1995hc,Magorrian:1998tg,Tremaine:2002qa}), and
are believed to play an important role in the evolution of
their host galaxies during an active phase in which they accrete
matter (e.g., \citealp{Ferrarese:2000ij,Gebhardt:2000bs,Kormendy:2013mj}). During such phases, they are
observed as Active Galactic Nuclei (AGN). 
In the most luminous AGN, accretion is likely triggered by major galaxy mergers (e.g., \citealp{Treister:2012fk}). An important stage in the
life-cycle of SMBHs is believed to happen during a dust-enshrouded
phase, when SMBHs accrete most of their mass, before blowing out the
material (e.g., \citealp{Martinez-Sansigre:2005lh}, \citealp{Glikman:2007ys}, \citealp{Urrutia:2008vn}, \citealp{LaMassa:2016ly}) and evolving into a blue unobscured source (e.g.,
\citealp{Hopkins:2006fv}). During this obscured phase the system is
expected to be extremely bright in the infrared (IR). The first
objects with these characteristics were discovered in large numbers by
the {\it Infrared Astronomical Satellite} ({\it IRAS}), and are called
luminous [$L_{\rm\,IR}(8-1000\,\mu\rm\,m)=10^{11}-10^{12}$
  $L_{\odot}$] and ultra-luminous ($L_{\rm\,IR}= 10^{12}-10^{13 }$ $L_{\odot}$)
infrared galaxies (LIRGs and ULIRGs, respectively; e.g., \citealp{Sanders:1996uq}, \citealp{Farrah:2003mi}, \citealp{Lonsdale:2006fu}, \citealp{Imanishi:2007pi}, \citealp{Veilleux:2009ff}). Subsequently, submillimeter galaxies (SMGs; e.g.,
\citealp{Blain:2002dz,Alexander:2005fu,Casey:2014kl}) at $z\sim 2-4$ were discovered
at longer wavelengths, while {\it Spitzer} surveys identified a
population of Dust-Obscured Galaxies at $z\sim 2$ (DOGs; e.g.,
\citealp{Yan:2007dq,Dey:2008cr,Fiore:2009nx}, see also \citealp{Toba:2015uq} and \citealp{Toba:2016fk} for studies of DOGs selected using other facilities).

More recently, the {\it Wide-field Infrared Survey Explorer} satellite
({\it WISE}, \citealp{Wright:2010fk}) has surveyed the whole sky in
four mid-infrared (mid-IR) bands, discovering new populations of
hyper-luminous ($L_{\rm\,IR} = 10^{13}-10^{14}$ $L_{\odot}$; e.g., \citealp{Eisenhardt:2012ve,Wu:2014,Hainline:2014wa}) and
extremely luminous ($L_{\rm\,IR} > 10^{14} L_{\odot}$; \citealp{Tsai:2015qf})
infrared galaxies (HyLIRGs and ELIRGs, respectively). This was accomplished by selecting objects that are
faint or undetected in the $W1$ (3.4\,$\mu$m) and $W2$ (4.6\,$\mu$m)
bands, but bright in the $W3$ (12\,$\mu$m) and $W4$ (22\,$\mu$m)
bands. Overall $\sim 1000$ of these sources were discovered across the
entire extragalactic sky (i.e., $\sim 1$ per 30 deg$^2$) \citep{Eisenhardt:2012ve}. Spectroscopic redshifts for $115$
``$W1W2$-dropouts'' are currently available \citep{Assef:2015zr}, and most of these objects are
at $z \gtrsim 1.5$, with the current highest redshift being $z = 4.601$
(\citealp{Tsai:2015qf,Diaz-Santos:2016ly}). These sources are typically optically faint, and their IR spectral energy distributions (SEDs) peak at rest-frame $\lambda\sim 20\,\mu$m, implying dust hotter ($T\gg 60$\,K) than in ULIRGs,
SMGs or DOGs. They are therefore referred to as Hot, Dust-Obscured Galaxies (Hot DOGs;
\citealp{Wu:2012bh}). It has been shown that for these ELIRGs the 1--20$\mu$m luminosity is always larger than the infrared luminosity above 20$\mu$m \citep{Tsai:2015qf}. The lack of a far-IR peak in their SEDs implies that the dominant energy sources are luminous heavily obscured AGNs
and not extreme starbursts \citep{Wu:2012bh,Tsai:2015qf}. {\it Hubble Space Telescope} and Keck/NIRC2 observations of Hot DOGs show strong lensing is unlikely \citep{Eisenhardt:2012ve,Wu:2014,Tsai:2015qf,Fan:2016sf}, while X-ray studies (\citealp{Stern:2014kx,Assef:2016qf,Piconcelli:2015uq}, this work) show that they contain very powerful AGN. The number density of Hot DOGs is comparable to that of type 1 AGN with similar luminosities at redshifts $2<z<4$ \citep{Stern:2014kx,Assef:2015zr}.
The most-luminous known galaxy in the Universe,
WISE\,J2246-0526 ($L_{\rm\,bol}=3.5\times 10^{14} L_{\odot}$, \citealp{Tsai:2015qf}), is a
Hot DOG. Recent ALMA observations of this object have found
evidence wide velocity spread, consistent with strong turbulence or isotropic outflows, which implies that the system is
blowing out its interstellar medium, and might be in the
process of becoming an unobscured quasar
\citep{Diaz-Santos:2016ly}. Hot DOGs might therefore represent a key phase in the evolution of AGN.

\begin{figure*}[t!]
\centering
\begin{minipage}[!b]{.48\textwidth}
\centering
\fbox{\includegraphics[width=8.5cm]{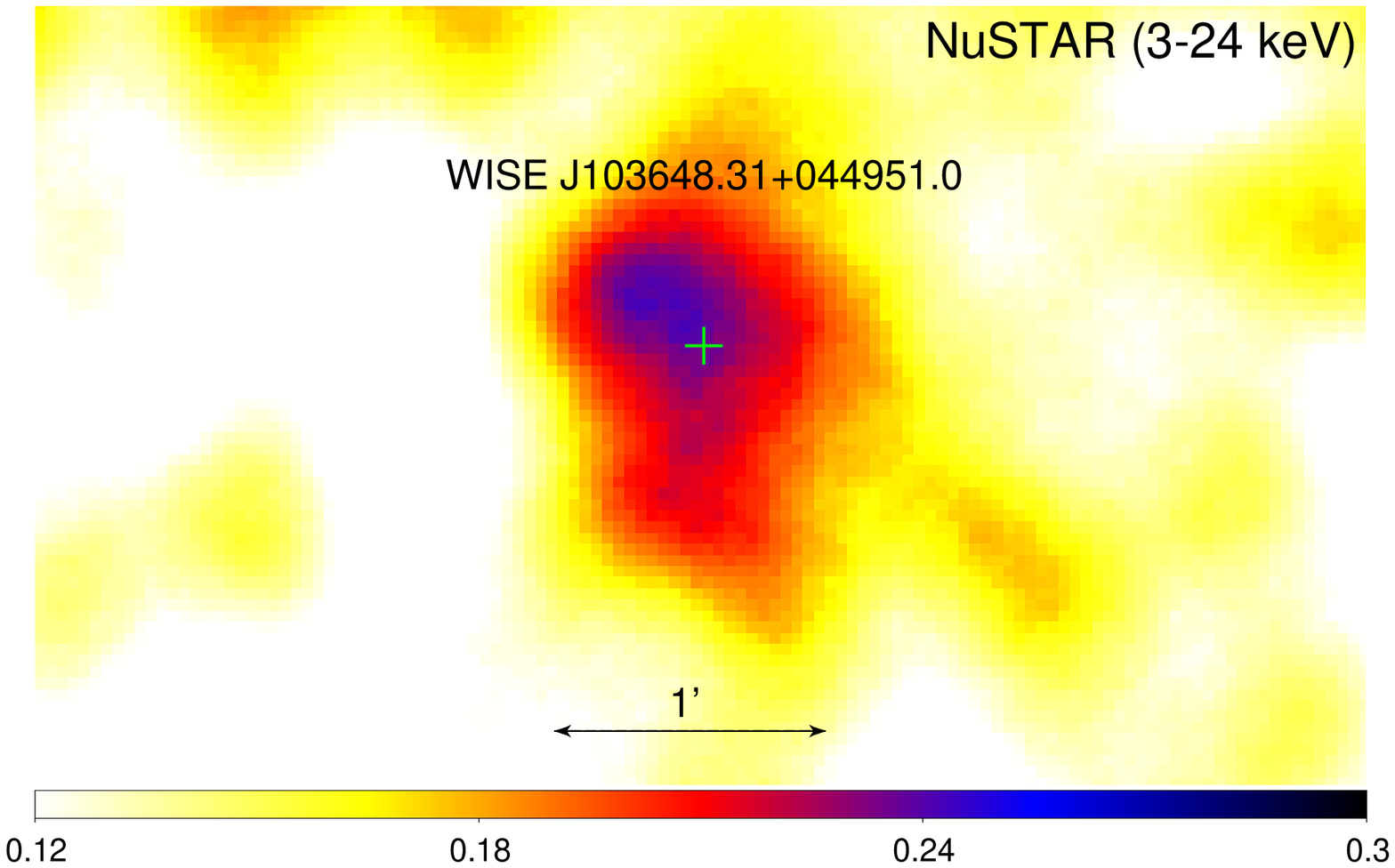}}\end{minipage}
\begin{minipage}[!b]{.48\textwidth}
\centering
\fbox{\includegraphics[width=8.5cm]{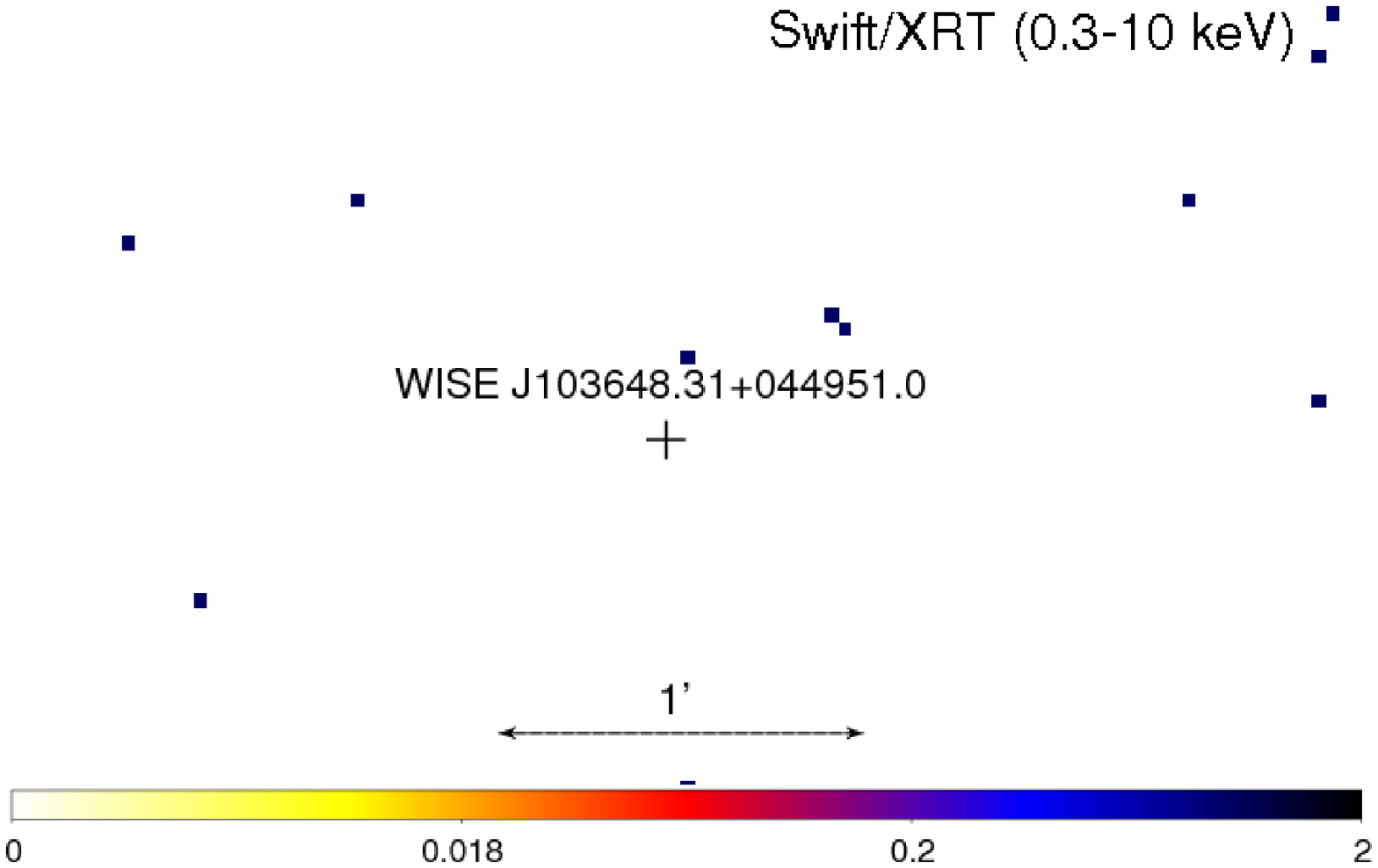}}\end{minipage}
\begin{minipage}[!b]{.48\textwidth}
\centering
\fbox{\includegraphics[width=8.5cm]{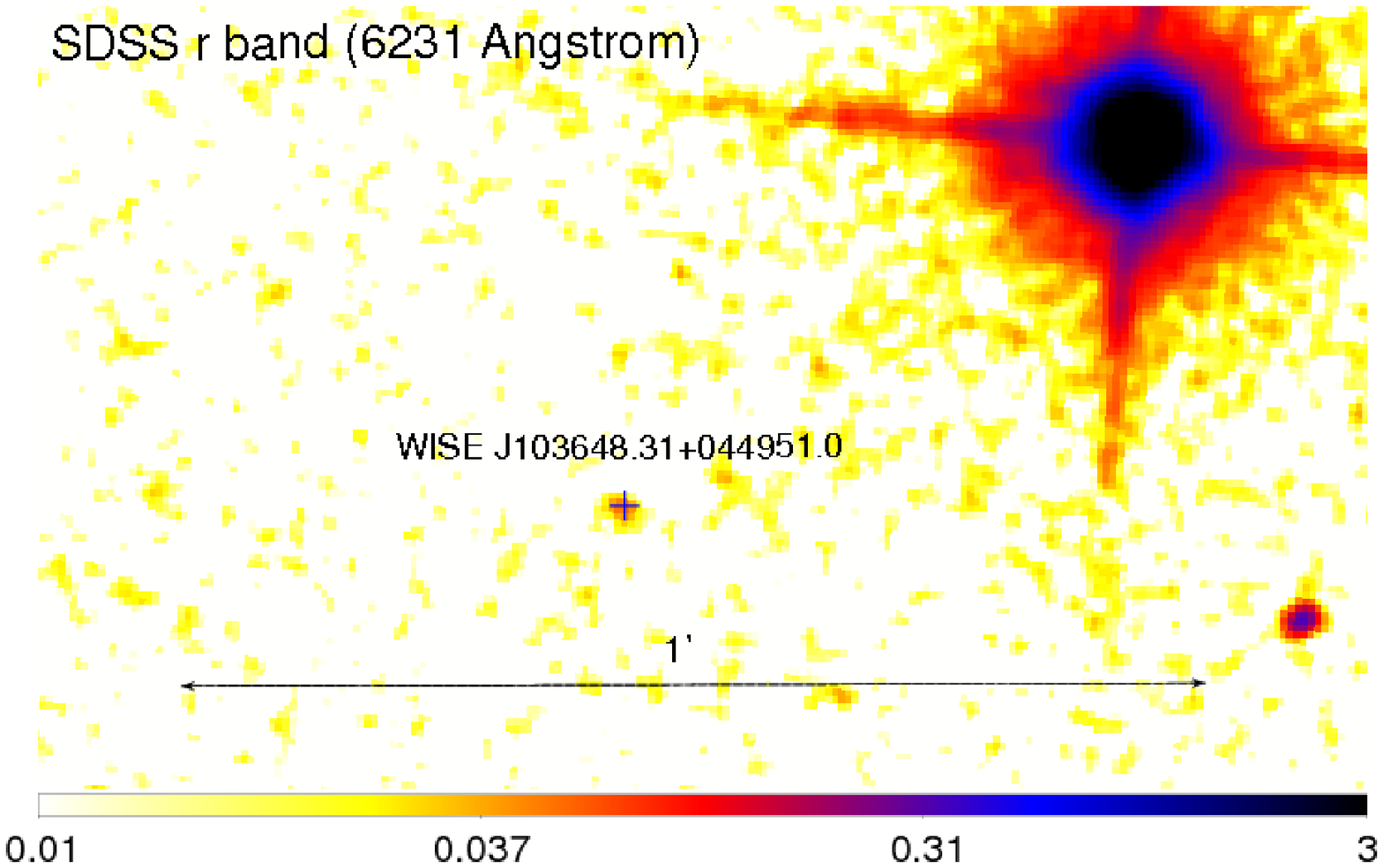}}\end{minipage}
\begin{minipage}[!b]{.48\textwidth}
\centering
\fbox{\includegraphics[width=8.5cm]{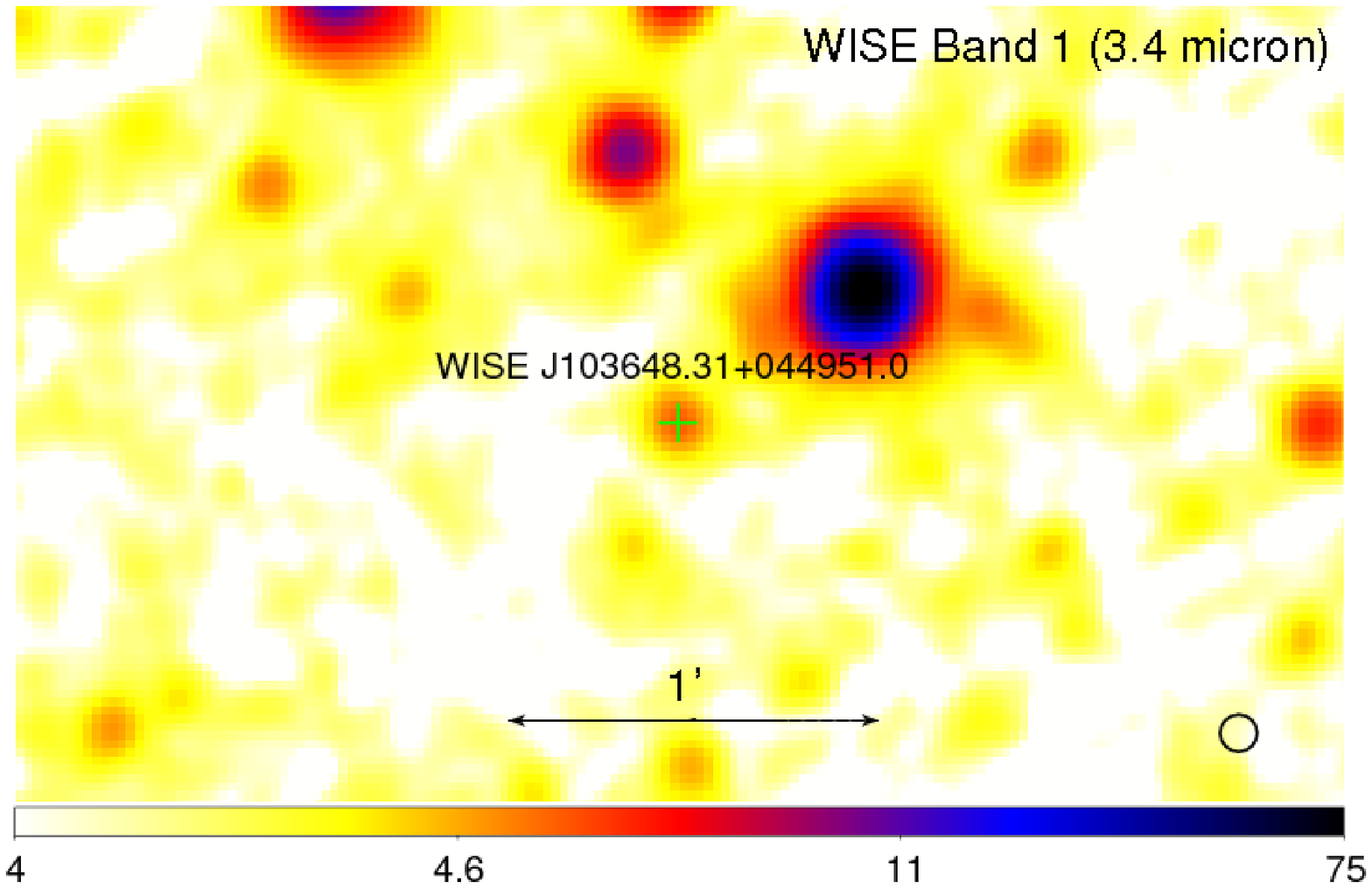}}\end{minipage}
\begin{minipage}[!b]{.48\textwidth}
\centering
\fbox{\includegraphics[width=8.5cm]{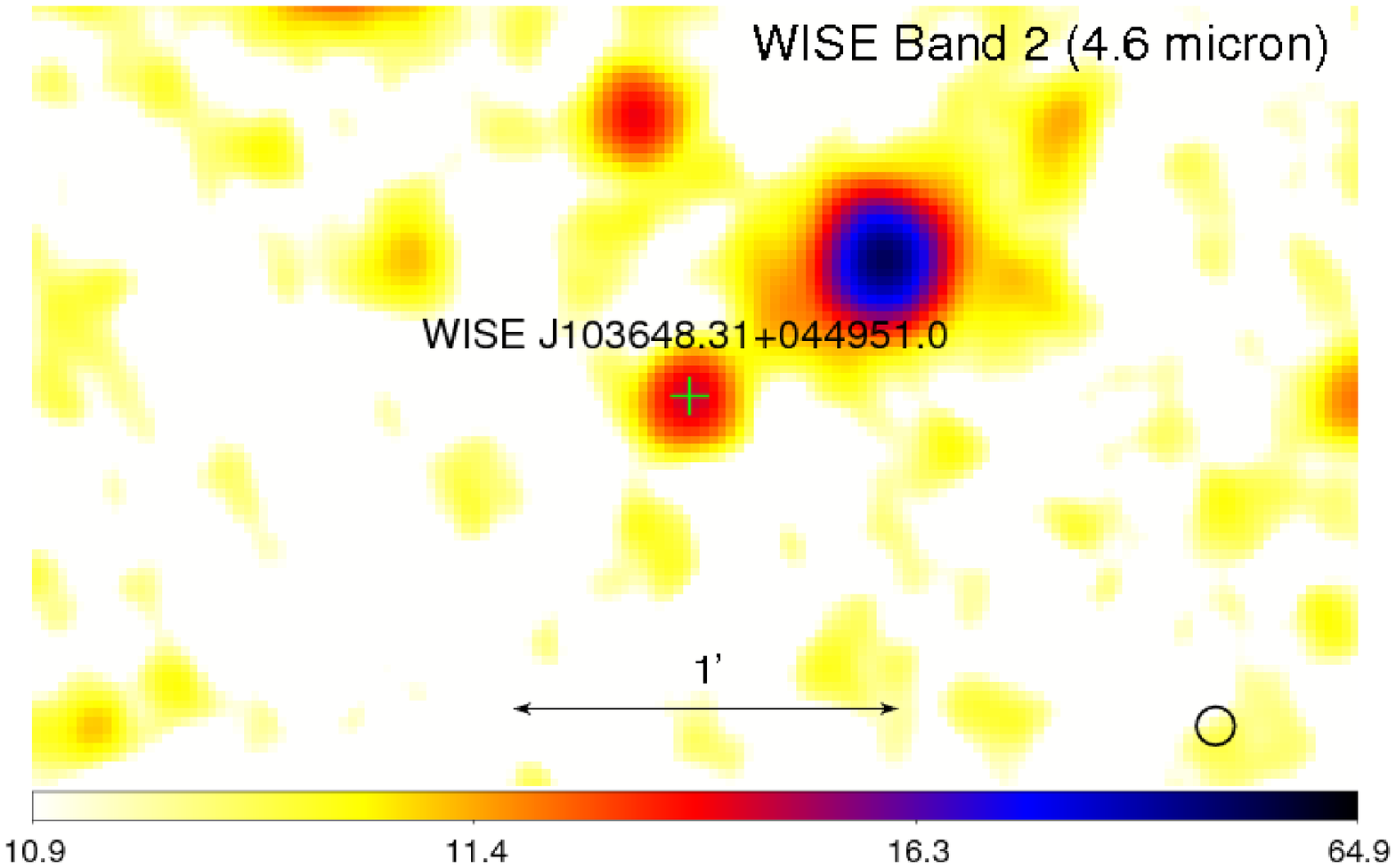}}\end{minipage}
\begin{minipage}[!b]{.48\textwidth}
\centering
\fbox{\includegraphics[width=8.5cm]{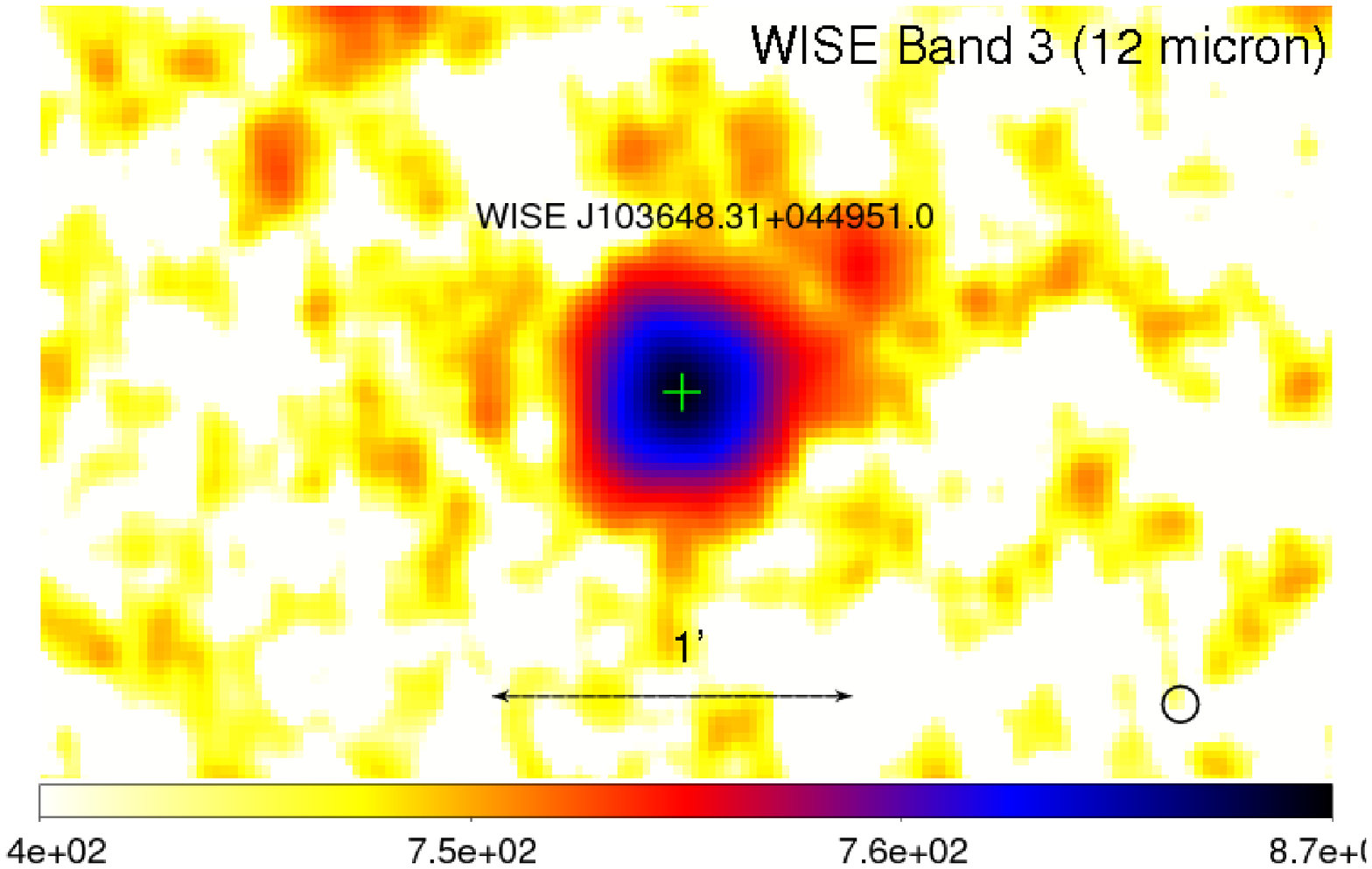}}\end{minipage}
\begin{minipage}[!b]{.48\textwidth}
\centering
\fbox{\includegraphics[width=8.5cm]{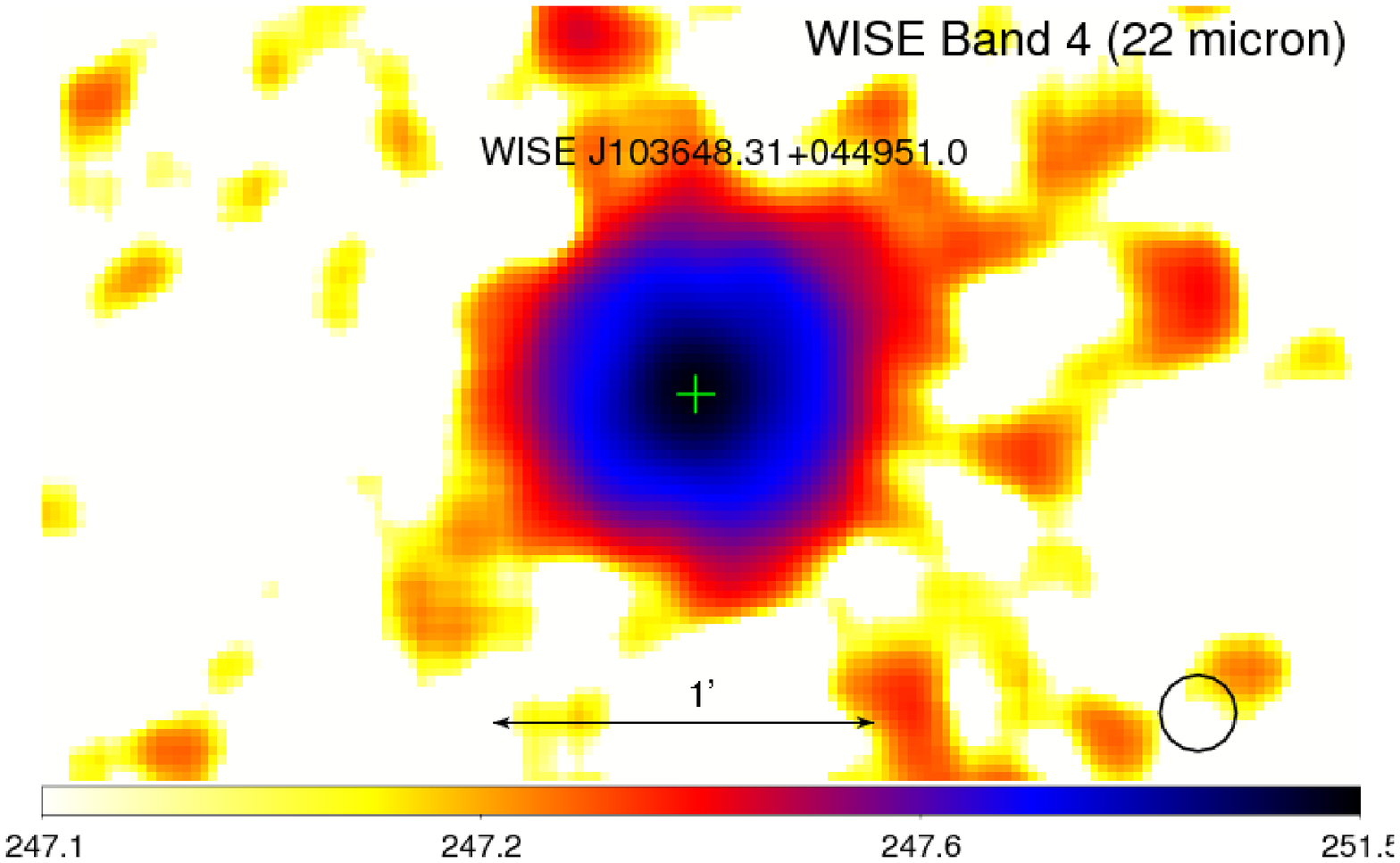}}\end{minipage}
\begin{minipage}[!b]{.48\textwidth}
\centering
\fbox{\includegraphics[width=8.5cm]{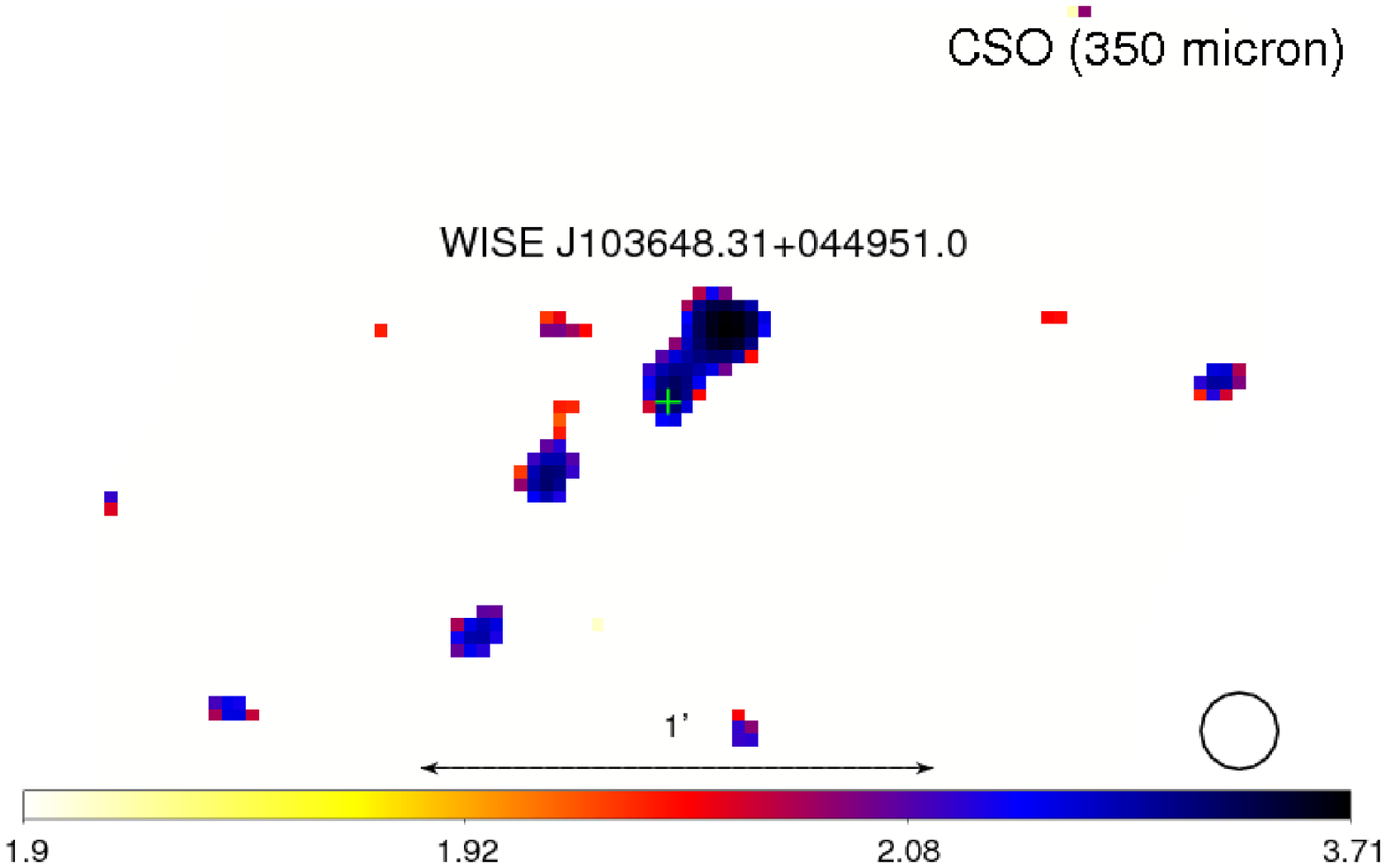}}\end{minipage}
 \begin{minipage}[t]{\textwidth}
  \caption{Images of the field around WISE\,J1036+0449 from
    the hard X-rays to the far-IR. From the top to the bottom, the
    panels show the images obtained by {\it NuSTAR} FPMA (3--24\,keV),
    {\it Swift}/XRT (0.3--10\,keV), SDSS $r$-band (6231\AA),
    {\it WISE} band 1 ($3.4\,\mu$m), band 2 ($4.6\,\mu$m), band 3
    ($12\,\mu$m) and band 4 ($22\,\mu$m), and by CSO ($350\,\mu$m). The {\it
      NuSTAR} image was obtained by combining the three observations, as
    described in Section\,\ref{sect:specAnalysis}, and was smoothed
    with a Gaussian kernel of radius 12 pixels. The SDSS image was
    smoothed with a Gaussian kernel of radius 2 pixels. The CSO image
    shows the signal-to-noise ratio per pixel. The crosses show the position of WISE\,J1036+0449, while the circles in the bottom right the size of the beam. In all images North is up and East is left.}
\label{fig:images}
 \end{minipage}
\end{figure*}

In AGN, much of the X-ray emission is produced in a compact region very close to
the SMBH ($\lesssim 10\,r_{\rm\,G}$; \citealp{Zoghbi:2012uq,De-Marco:2013fk} --- where $r_{\rm\,G}=G\,M_{\rm\,BH}/c^2$ is the gravitational radius of the SMBH). X-ray observations are therefore a potent tool to infer the line-of-sight column density to the central engine ($N_{\rm\,H}$). The relation between the bolometric and X-ray output of Hot DOGs also sheds light on the physical conditions of the X-ray emitting plasma. Hot DOGs are therefore  excellent laboratories
for probing the structure of the accretion flow at the highest luminosities, although they are not yet well-studied in the X-ray band,
with only a handful having been observed by X-ray
facilities to date. \citet{Stern:2014kx} reported on two Hot DOGs observed with
{\it NuSTAR} and {\it XMM-Newton}, plus an additional
source observed only by {\it XMM-Newton}. All three
targets are at $z\sim 2$. Neither target observed by {\it NuSTAR}
yielded a significant detection, while two of the three objects were faintly
detected by {\it XMM-Newton}, implying that the sources are
either X-ray weak or heavily obscured by column densities $N_{\rm
  H}\gg 10^{24}~\rm cm^{-2}$. Similar results were obtained by
\citet{Piconcelli:2015uq}, who studied a 40~ks {\it{XMM-Newton}}
spectrum of WISE\,J1835+4355, a Hot DOG at $z= 2.298$, and found $N_{\rm H}\gg 10^{23}~\rm cm^{-2}$, with the source likely being reflection dominated (Zappacosta et al., in prep.). Recently, \citet{Assef:2016qf}
found evidence of similar levels of obscuration in the X-rays for another Hot DOG, WISE\,J0204--0506 ($z=
2.100$), using a serendipitous off-axis {\it Chandra} observation (160~ks exposure). 

\begin{figure*}
  \begin{center}
      \epsscale{0.75}
    \plotone{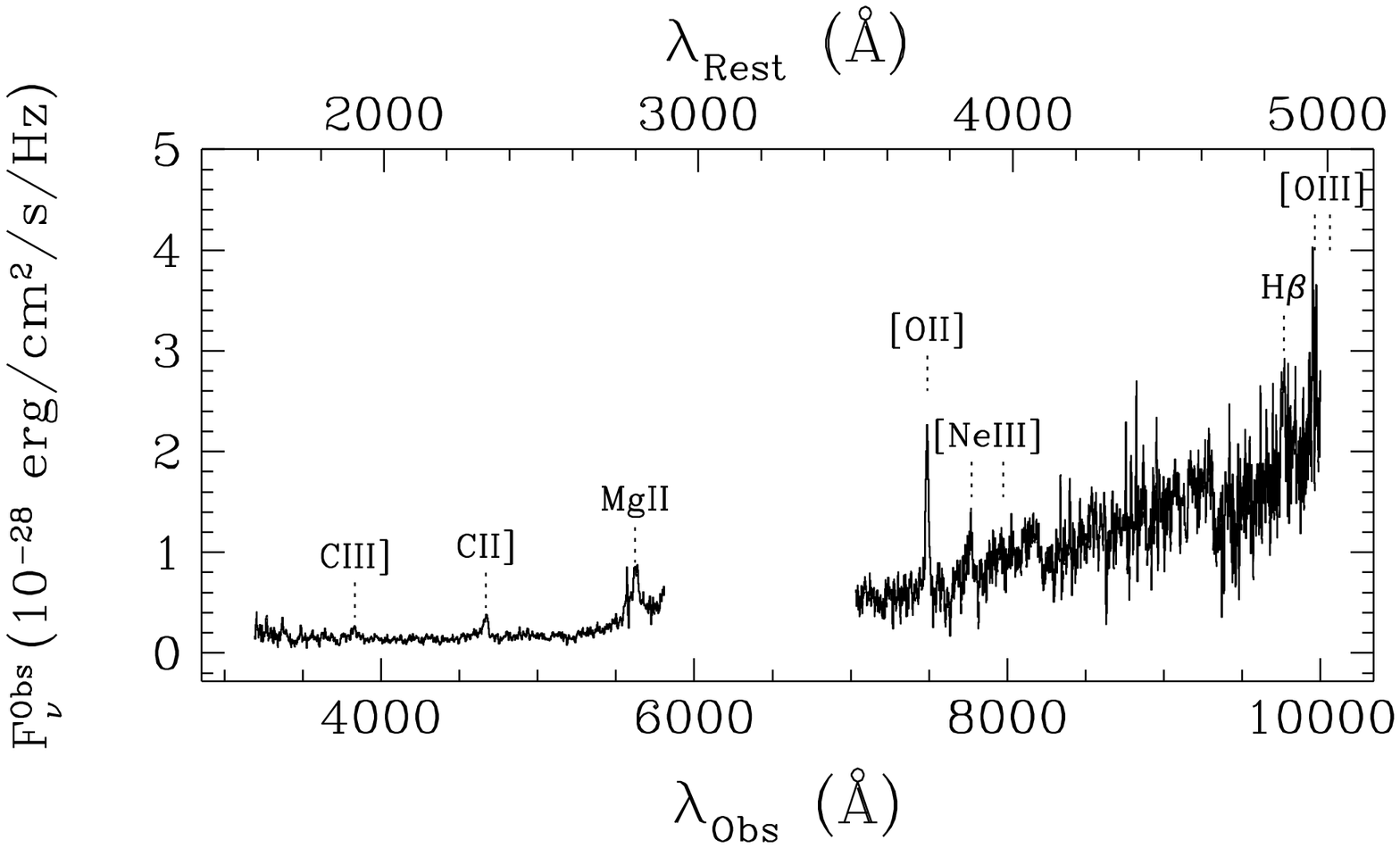}
    \caption{UV/optical spectrum of \sname, obtained with the LRIS
      instrument at the Keck Observatory. See $\S$\ref{sect:Opticalspec} for details.}
    \label{fg:lris_spec}
  \end{center}
\end{figure*}

To constrain better the X-ray absorption of Hot DOGs,
and hence their intrinsic X-ray luminosities, it is necessary to obtain
reliable detections at $E\gtrsim 10\rm\,keV$, where their emission is
less affected by neutral gas absorption (e.g., \citealp{Lansbury:2015dq}, \citealp{Annuar:2015qf}, \citealp{Puccetti:2016zr}, \citealp{Tanimoto:2016vn}, \citealp{Ricci:2016kq,Ricci:2016fk}). The simplest way to do this
is to observe brighter, lower-redshift sources. However,
\citet{Assef:2015zr} show that the number of Hot DOGs at such redshift is
very small, in part due to an inherent bias in their selection function,
with fast space density evolution also a likely contributing factor. Furthermore, Hot DOGs that happen to be at lower redshifts are biased toward being much less luminous than their higher redshift counterparts due, at least in part, to the strict requirements of the selection function on the $W1$ flux. 

A new selection technique, as discussed in the following section, allows identification of a significant
population of Hot DOGs at $z\sim 1$ (Assef et al. in prep.). We report here on the study of
one of these new objects, WISE\,J103648.31+044951.0 (WISE\,J1036+0449). In this paper, we show that the SED of WISE\,J1036+0449 at z=1.009 peaks in the mid IR, similarly to Hot DOGs at higher redshift. Exploiting three {\it NuSTAR}
observations, we are able to constrain the line-of-sight column
density and its intrinsic X-ray luminosity. WISE\,J1036+0449 is one of
the closest Hot DOGs known ($z=1.009$) and, given its relative
proximity, it could become an important case study of this
interesting population of AGN.

The paper is structured as follows. In $\S$\ref{sect:selection} we describe the selection method, and show that the SED of WISE\,J1036+0449 is consistent with those of other Hot DOGs. In $\S$\ref{Sect:data_analysis} and $\S$\ref{sect:specAnalysis} we report on the X-ray observations available and on the X-ray spectral analysis, respectively. In $\S$\ref{sect:discussion} we discuss the possible intrinsic X-ray weakness of Hot DOGs and their absorption properties, while in $\S$\ref{sect:conclusion} we report our conclusions. Throughout the paper we use Vega magnitudes and adopt standard cosmological parameters ($H_{0}=70\rm\,km\,s^{-1}\,Mpc^{-1}$, $\Omega_{\mathrm{m}}=0.3$, $\Omega_{\Lambda}=0.7$). Unless otherwise stated, all uncertainties are quoted at the 90\% confidence level.

\section{Source selection and Spectral Energy Distribution}\label{sect:selection}

\begin{figure*}
  \begin{center}
\includegraphics[width=0.7\textwidth,angle=0]{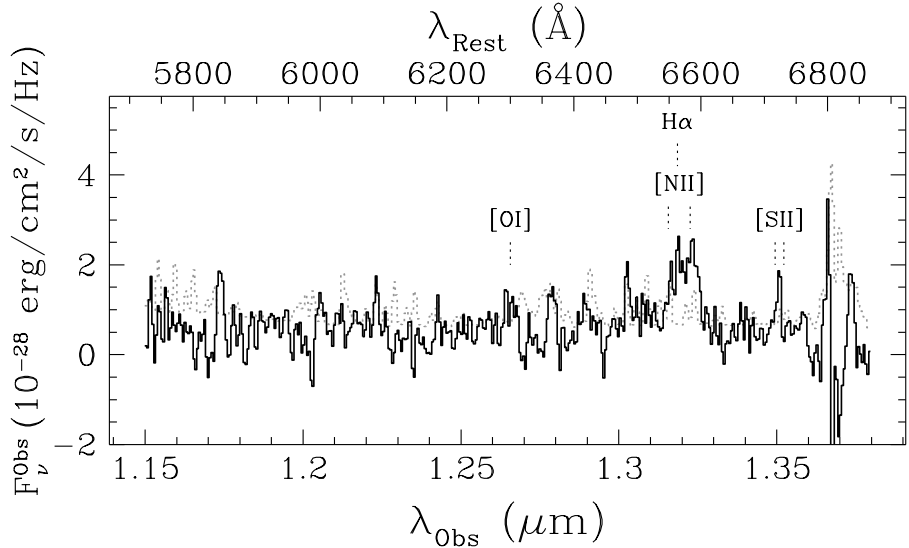}
    \caption{Near-IR Keck/NIRSPEC spectrum of WISE\,J1036+0449 (see $\S$\ref{sect:NIRspec} for details). The dashed line shows the error spectrum.}
    \label{fig:NIRspec}
  \end{center}
\end{figure*}

\citet{Assef:2015zr} presented the redshift distribution of a large
sample of Hot DOG candidates, showing that it is
bimodal, with almost all confirmed Hot DOGs at $z\sim 2-4$, and a small number
of apparent contaminant sources at $z\lesssim 0.5$. There is a significant
dearth of Hot DOGs in the $z\sim 1-2$ redshift
range, and \citet{Assef:2015zr} argue that while a strong redshift
evolution toward this redshift range is likely, the results could be in
part driven by an inherent bias against $z\lesssim 2$ objects in their
selection function. As discussed in detail by
\citet{Eisenhardt:2012ve}, Hot DOGs are selected purely based on their
{\it WISE} magnitudes (specifically those from the All-Sky Data
Release\footnote{\url{http://wise2.ipac.caltech.edu/docs/release/allsky/}}),
without the use of supporting observations.  Hot DOGs are
required to have $W1>$17.4\,mag and satisfy either that i) $W3<$10.6\,mag and
$W2-W3>$5.3\,mag, or that ii) $W4<$7.7\,mag and $W2-W4>$8.2\,mag. The requirement
of faintness in the $W1$ and $W2$ bands strongly biases the sample against
the most-luminous $z\sim 1$ objects, where $W1$ would sample the rest-frame 1.6$\,\mu$m
maximum of the host-galaxy stellar emission. Such biases must be
considered when studying the redshift and luminosity evolution of this
population.

To mitigate these biases, we have devised a complementary {\it WISE} color selection
function which allows to target Hot DOGs at $z\sim1-2$. Specifically, we select objects whose All-Sky Data Release
magnitudes meet all the following three requirements: (i) $W1<$17.4\,mag,
(ii) $W1-W3>$7\,mag, and (iii) $W1-W4>$10\,mag. The color requirements were
chosen to specifically select objects in the $z\sim 1-2$ range with
IR SEDs similar to that of WISE\,J1814+3412 \citep{Eisenhardt:2012ve},
which is one of the best-studied Hot DOGs in the literature. Requirement (i) makes the selection function
complementary to that of $z>1.5$ Hot DOGs (which have $W1>$17.4\,mag), and reflects the fact that deeper {\it Spitzer} IRAC detections of z $>$ 1.5 Hot DOGs (Assef et al. 2015, Tsai et al. 2015) implies that z$\sim$1 sources with similar SEDs will be detected by {\it WISE} at this level. There are only 153 objects in the
entire extragalactic sky that meet these criteria, and an extensive
follow-up campaign is underway to assess the completeness and
contamination of this selection. These results will be
reported in detail elsewhere \citep{Assef:2016hp}.

One of the first objects selected by the criteria defined above is
WISE\,J1036+0449  (Fig.\,\ref{fig:images}), which would not meet the Hot DOGs selection
criteria because it is too bright in the W1 band. In the next sections we discuss some of the follow-up
observations carried out to demonstrate that it is a bona fide Hot
DOG in the targeted redshift range.

\subsection{Spectroscopic Follow-Up of \sname}

\subsubsection{Optical spectroscopy}\label{sect:Opticalspec}

Optical spectroscopy for \sname\ was obtained on UT 2012 April 20
using the Low Resolution Imaging Spectrometer (LRIS, \citealp{Oke:1995fk}) at the Keck
observatory with a slit width of 1.5\,arcsec. The data were reduced using the standard
IRAF\footnote{\url{http://iraf.noao.edu/}} tools. The spectrum (Fig.\,\ref{fg:lris_spec}) shows a red continuum and several well
detected narrow emission lines. From blue to red, these are C\,{\sc
  iii}]$\lambda$1909\AA, C\,{\sc ii}]$\lambda$2324\AA, Mg\,{\sc
      ii}$\lambda$2798\AA, [O\,{\sc ii}]$\lambda$3727\AA, [Ne\,{\sc
        iii}]$\lambda$3869\AA, H$\beta$ and [O\,{\sc
        iii}]$\lambda$4959\AA. Using these emission lines, we estimate
    a redshift of $z=1.009\pm0.002$, fully consistent with the redshift range expected.

\subsubsection{Near-infrared spectroscopy}\label{sect:NIRspec}

We also obtained near-IR spectroscopy for WISE\,J1036+0449 on UT 2016 January 27 using the NIRSPEC instrument \citep{McLean:1998kq} at the Keck Observatory. Observations were carried out in the $J$-band using the 42\arcsec\ $\times$0.57\arcsec\ longslit in the low-resolution mode, and the source was observed for a total exposure time of 4$\times$500 s in an ABBA sequence. The observations were reduced using a combination of the IRAF NIRSPEC tools provided by the Keck Observatory \footnote{\url{ftp://ftp.keck.hawaii.edu/pub/ObservingTools/iraf/keck.tar.gz}} and standard IRAF tools. The wavelength calibration was done using the sky emission lines.

\subsubsection{Optical and Near-IR spectral analysis}\label{sect:NIRoptSpec}

The emission features observed in the optical spectrum, in particular the presence of C\,{\sc iii}]
and Mg\,{\sc ii}, strongly imply that \sname\ is an obscured AGN (e.g., \citealp{Stern:1999uq}). Such
spectral characteristics are commonly observed in Hot DOGs
\citep{Wu:2012bh,Assef:2016qf,Eisenhardt:2016zc}, supporting the idea that
\sname\ belongs to this class of objects.

The Mg\,{\sc ii} emission line has a significantly broad base,
implying that part of the broad-line region emission might be visible despite the
large amount of obscuration towards the accretion disk. Assuming that Mg\,{\sc ii} is broadened by the gravitational potential of the SMBH, and using the \cite{McGill:2008bh} calibration, we find that the black hole mass is $M_{\rm\,BH}\simeq 2 \times 10^8 M_{\odot}$ (see Wu et al. in prep. for details of the interpretation of broad lines in Hot DOGs). Such a broad base is not visible, however, for the H$\beta$ emission line. In the following we discuss the near-IR spectrum of the source, and show that for H$\alpha$, the signal-to-noise ratio is too low to definitively measure the line-width but it might be broad enough to warrant an intermediate AGN
classification for this object. We caution, however, that such an intermediate type might not be due to a lower obscuration, but to reflected light from the central engine as in the Hot DOG studied by \cite{Assef:2016qf}. We note
that the [O\,{\sc ii}] doublet has an observed-frame width of
approximately 20\AA, which is wider than the spectral resolution of
$\sim 10$\AA\ of the observations. No asymmetry is observed, as could
be expected for quasar outflows. The excess width of $\sim
17\,$\AA\ implies a FWHM of approximately $700~\rm km~\rm s^{-1}$. Such
width is consistent with what would be expected for emission
associated with the narrow-line region of a quasar. However, in less-extreme galaxies, [O\,{\sc
    ii}] is typically related to star-formation in the host
galaxy \citep[see, e.g.][]{Kauffmann:2003kk}, implying that it could
be due to turbulence in the inter-stellar medium. Recently,
\citet{Diaz-Santos:2016ly} used ALMA to study the [C\,{\sc
    ii}]$\lambda$157.7$\,\mu$m emission line in the Hot DOG W2246--0526
at $z=4.601$ \citep{Tsai:2015qf}. \citet{Diaz-Santos:2016ly} found a FWHM of
$600~\rm km~\rm s^{-1}$ for this emission line and showed that this
was consistent with strong, [CII]-emission-region-wide turbulence in the interstellar medium (ISM) of the galaxy. 

The near-IR spectrum (Fig.\,\ref{fig:NIRspec}) shows a clear detection of the H$\alpha$ emission line as well as of the [N\,{\sc ii}]$\lambda\lambda$6549,6583\AA\ and [S\,{\sc ii}]$\lambda\lambda$6717,6731\AA\ doublets, although the $S/N$ of the spectra is too low to accurately measure line-widths as was done in the optical. The observed features are consistent with what would be expected for a type 2 AGN, as implied by the optical spectrum. The emission lines show a small systematic offset to the red, suggesting a slightly larger redshift closer to $z=1.010$, which is consistent with the optical value within the uncertainties. The continuum level at the blue end of the near-IR spectrum seems to be a factor of $\sim$4 below that at the red end of the optical spectrum. While no effort has been made to take into account the variations of the seeing between the science and calibration targets, which could have resulted in biases to the absolute level of the flux calibration, such a large difference could imply that the host galaxy is significantly more extended than the slit-width of 0.57\arcsec (4.6~kpc) in the $J$ band.

\begin{figure}
  \begin{center}
\includegraphics[width=0.5\textwidth,angle=0]{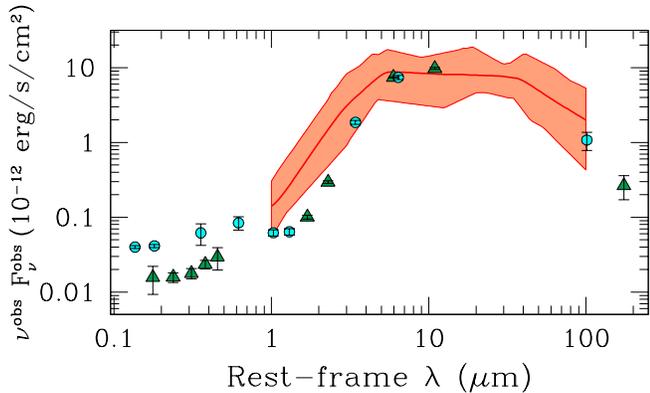}
    \caption{Rest-frame UV through far-IR broad-band
      SED of WISE\,J1036+0449 (green triangles). The green triangles show the flux measured in the
      SDSS $u^{\prime}r^{\prime}g^{\prime}i^{\prime}z^{\prime}$ bands,
      the four {\it WISE} photometric bands, and in the CSO/SHARC II
      350$\,\mu$m band. The red solid line shows the geometric mean SED of the
      sources studied by \citet{Tsai:2015qf} using {\it WISE} and
      {\it{Herschel Space Observatory}} observations, scaled to the
      {\it WISE} photometry of \sname. The cyan circles show the SED of WISE\,J1814+3412 (renormalized to a similar flux level of WISE\,J1036+0449 for comparison), while the light red region shows the range
      covered by all the sources in the study of \citet{Tsai:2015qf}. The SED is shown in $\nu F_{\nu}$ for comparison with \citet{Tsai:2015qf}.}
    \label{fig:SED}
  \end{center}
\end{figure}

\subsection{UV through far-IR SED of \sname}

One of the main characteristics of Hot DOGs is their distinctive multi-wavelength SED
\citep[see][]{Eisenhardt:2012ve,Wu:2012bh,Jones:2014jj,Assef:2015zr,Tsai:2015qf}. In
these objects the rest-frame mid- through far-IR is dominated by emission from a highly obscured, hyper-luminous AGN, with little to no
contribution from a cold dust component that would be traditionally
associated with star-formation in the host galaxy. The optical, on the
other hand, is typically dominated by the emission from the much less-luminous host galaxy. The SED peaks at a rest-frame
wavelength of $\sim 20\,\mu\rm m$ and, hence, shows extremely red
colors between the optical/near-IR and the mid-IR, but very blue colors
between the mid- and far-IR.

\subsubsection{Data and comparison with the typical SED of Hot DOGs}

The UV through far-IR SED of \sname\ is shown in Fig.\,\ref{fig:SED} and the photometry is listed in Table\,\ref{tab:mwphotometry}. The figure shows archival optical photometry in the
$u^{\prime}r^{\prime}g^{\prime}i^{\prime}z^{\prime}$ bands from the
Sloan Digital Sky Survey Data Release 12 \citep{Alam:2015aa} as well as
the mid-IR photometry in the four {\it WISE} bands from the AllWISE data
release \citep{Cutri:2012lq}. The {\it WISE} images, as well as the SDSS $r^{\prime}$ image, are
shown in Fig.\,\ref{fig:images}. We also obtained additional
photometry for this target at 350\,$\mu$m with the SHARC II instrument
\citep{Dowell:2003sh} at the Caltech Submillimeter Observatory
(CSO). We observed the target for 50 minutes on UT 2012 March 21 and
for 40 minutes on UT 2012 December 15. The data are shown in Fig.\,\ref{fig:images}. The observations were analyzed in the same manner as
in \citet{Wu:2012bh}, and a total flux density of $31\pm 11~\rm mJy$
was measured for the source\footnote{As a calibrator we used IRC\,+10216, and we adopted a calibration uncertainty of 20\%.}. The photometry is shown in Fig.\,\ref{fig:SED} and reported in Table\,\ref{tab:mwphotometry}.

\begin{table}
\begin{center}
\caption[]{Multi-wavelength photometry.}
\label{tab:mwphotometry}
\begin{tabular}{ccc}
\noalign{\smallskip}
\hline \hline \noalign{\smallskip}
(1) & \multicolumn{1}{c}{(2)} & (3) \\
\noalign{\smallskip}
$\lambda_{\rm\,Rest}$ ($\mu$m) & $F_{\nu}$ (Jy) & $\Delta F_{\nu}$ (Jy) \\
\noalign{\smallskip}
\hline \noalign{\smallskip}
0.176	 &  $1.85\times10^{-6}$  & $  7.51\times10^{-7}  $   \\
\noalign{\smallskip}
 0.237 	&  $2.51\times10^{-6}$  & $ 3.70\times10^{-7}   $ \\   
\noalign{\smallskip}
 0.310 	&  $3.70\times10^{-6}$  & $ 5.79\times10^{-7}   $ \\   
\noalign{\smallskip}
 0.380 	&  $5.97\times10^{-6}$  & $ 7.70\times10^{-7}   $ \\   
\noalign{\smallskip}
 0.455 	&  $8.95\times10^{-6}$  & $ 2.97\times10^{-6}   $ \\   
\noalign{\smallskip}
 1.692 	&  $1.14\times10^{-4}$  & $ 6.59\times10^{-6}   $ \\   
\noalign{\smallskip}
 2.290 	&  $4.53\times10^{-4}$  & $ 1.79\times10^{-5}   $ \\   
\noalign{\smallskip}
 5.973 	&  $2.96\times10^{-2}$  & $ 5.46\times10^{-4}   $ \\   
\noalign{\smallskip}
10.951 	&  $7.17\times10^{-2}$  & $ 2.05\times10^{-3}   $ \\   
\noalign{\smallskip}
174.22 	&  $3.10\times10^{-2}$  & $ 1.10\times10^{-2}   $ \\  
\noalign{\smallskip}
\noalign{\smallskip}
\hline
\noalign{\smallskip}
\end{tabular}
\tablecomments{The columns report (1) the rest-frame wavelength, (2) the flux and (3) the error on the flux.}
\end{center}
\end{table}

Figure\,\ref{fig:SED} also shows the SED of WISE\,J1814+3412 and the mean IR SED of the {20 Hot DOGs studied by
\citet{Tsai:2015qf}, which include the most-luminous ones known to
date with $L_{\rm IR}>10^{14}~L_{\odot}$ and cataloged as ELIRGs (none of which has so far been studied in the X-ray band). These objects are at
significantly higher redshifts than \sname, spanning the range between
$z=2.668$ and 4.601, with an average of $\langle z \rangle =
3.3$. \citet{Tsai:2015qf} presented observations obtained with the {\it Spitzer} IRAC instrument and with the {\it Herschel Space Observatory} PACS and SPIRE instruments in the 3.6, 4.5, 70, 160, 250, 350 and 500\,$\mu$m bands. The SED shape has been
obtained as the average of the fluxes of each source,
interpolated through power-laws in the rest-frame 1\,$\mu$m to 100\,$\mu$m
wavelength range (with some slight extrapolation where needed). The
mean SED was fitted to the {\it WISE} photometry of \sname\ for display
purposes.

The SED shape of \sname\ is  very similar to the SED of WISE\,J1814+3412 and qualitatively consistent with the mean
SED of the \citet{Tsai:2015qf} sources. 
The far-IR observations obtained with CSO show that \sname\
has a steeply dropping SED from rest-frame $\sim$10$\mu$m to
$\sim$100$\mu$m, which is one of the defining characteristics of the
Hot DOG population. This implies that cold-dust emission from star formation contributes
negligibly to the integrated SED, as such a component would be expected
to peak at $\sim$100$\mu$m (see discussions in \citealp{Wu:2012bh}, \citealp{Jones:2014jj}, \citealp{Tsai:2015qf}). Furthermore, the SED steadily drops between
rest-frame $\sim 5\,\mu\rm m$ and $\sim 1\,\mu\rm m$, which is also, by selection, a
defining characteristic of Hot DOGs and most IR-luminous galaxies. The mid-IR {\it WISE} colours are, in fact, somewhat redder than those of the objects studied
by \citet{Tsai:2015qf}, implying that they may be subject to even
larger extinction \citep[see][]{Assef:2015zr}.

The SED and the optical spectrum presented in
\S\ref{sect:Opticalspec} show that, although targeted with a
different selection function, \sname\ is a {\it{bona-fide}} Hot
DOG. Due to its lower redshift and significantly higher flux ($W4=5.18$\,mag) with respect to other Hot DOGs, \sname\ is an ideal target for hard
X-ray observations, which are able to probe the emission from the highly
obscured AGN (e.g., \citealp{Balokovic:2014dq}, \citealp{Gandhi:2014bh,Gandhi:2015uq}, \citealp{Arevalo:2014nx}, \citealp{Bauer:2015ve}, \citealp{Koss:2015bh}). 

\begin{table}
\begin{center}
\caption[]{X-ray Observations Log.}
\label{tab:obslog}
\begin{tabular}{clccc}
\noalign{\smallskip}
\hline \hline \noalign{\smallskip}
(1) & \multicolumn{1}{c}{(2)} & (3) & (4) & (5)\\
\noalign{\smallskip}
\hline \noalign{\smallskip}
1 &   {\it NuSTAR}													&  2014-11-01 10:01:07		& 60001156002	& 11.2		\\  
\noalign{\smallskip}
 1 &  {\it Swift}/XRT 	 											& 2014-11-01 23:08:58		& 00080818001 &	1.0		\\ 
\noalign{\smallskip}
\noalign{\smallskip}
\noalign{\smallskip}
 2 &  {\it NuSTAR}													& 2014-11-02 00:11:07  		&  60001156004 	& 36.1		\\ 
\noalign{\smallskip}
 2 &  {\it Swift}/XRT 	 											& 2014-11-02 00:46:58		& 00080818002 &	1.0		\\ 
\noalign{\smallskip}
\noalign{\smallskip}
\noalign{\smallskip}
3 &   {\it NuSTAR}													& 2014-12-22 20:21:07	 & 60001156006 	&  21.2		\\ 
\noalign{\smallskip}
3 &   {\it Swift}/XRT 	 											&  2014-12-22 22:05:59	& 00080818003	& 1.9		\\  
\noalign{\smallskip}
\noalign{\smallskip}
\hline
\noalign{\smallskip}
\end{tabular}
\tablecomments{The columns report (1) the observation number, (2) the X-ray facility, (3) the UT observation date, (4) the observation ID, and (5) the exposure time (in ks).}
\end{center}
\end{table}

\subsubsection{SED fitting}

The fact that the Mid-IR SED emission is dominated by radiation produced in an accreting SMBH is confirmed by fitting the data with the AGN template of \cite{Mullaney:2011lq}, following the procedure outlined in \cite{Stanley:2015gf}.  To reproduce the emission with a physical model we fit the rest-frame SED of WISE~J1036+0449 with a combination of AGN torus (e.g., \citealp{Schartmann:2008rm,Honig:2010jk,Stalevski:2012kq,Stalevski:2016ly}) and modified blackbody (BB) SEDs. The torus component is
supplied by the widely used \C\ models\footnote{\url{www.clumpy.org}},
introduced in \citet{Nenkova+2002,Nenkova+2008a,Nenkova+2008b}, and
comprises a collection of dusty clouds, each with optical depth
$\tau_V>1$ in the visual band (0.55\mic). A detailed description of this model can be found in Appendix\,\ref{sect:SEDmodel}.

The SED fitting method is described in Appendix\,\ref{sect:SEDfittApproach}. In the upper panels of Figure \ref{fig:fitting} we illustrate the best-fit SED
(maximum-a-posteriori likelihood model, MAP), i.e. the minimum reduced
$\chi^2$ composite model (torus + BB). In the left upper panel all the
available data points were fitted, while the right panel only
considers the IR data (1.6\mic\ rest-frame and beyond). The need to
run a fit without optical/UV data arises since these wavelength regions are possibly contaminated by the host galaxy contribution. This is the case for most Hot DOGs, as can be seen in \cite{Assef:2015zr} and \cite{Assef:2016hp}. The confidence contours obtained are shown in Fig.\,\ref{fig:contours_clumpy}.

%
\begin{figure*}
  \center
  \includegraphics[height=11.2cm,angle=90]{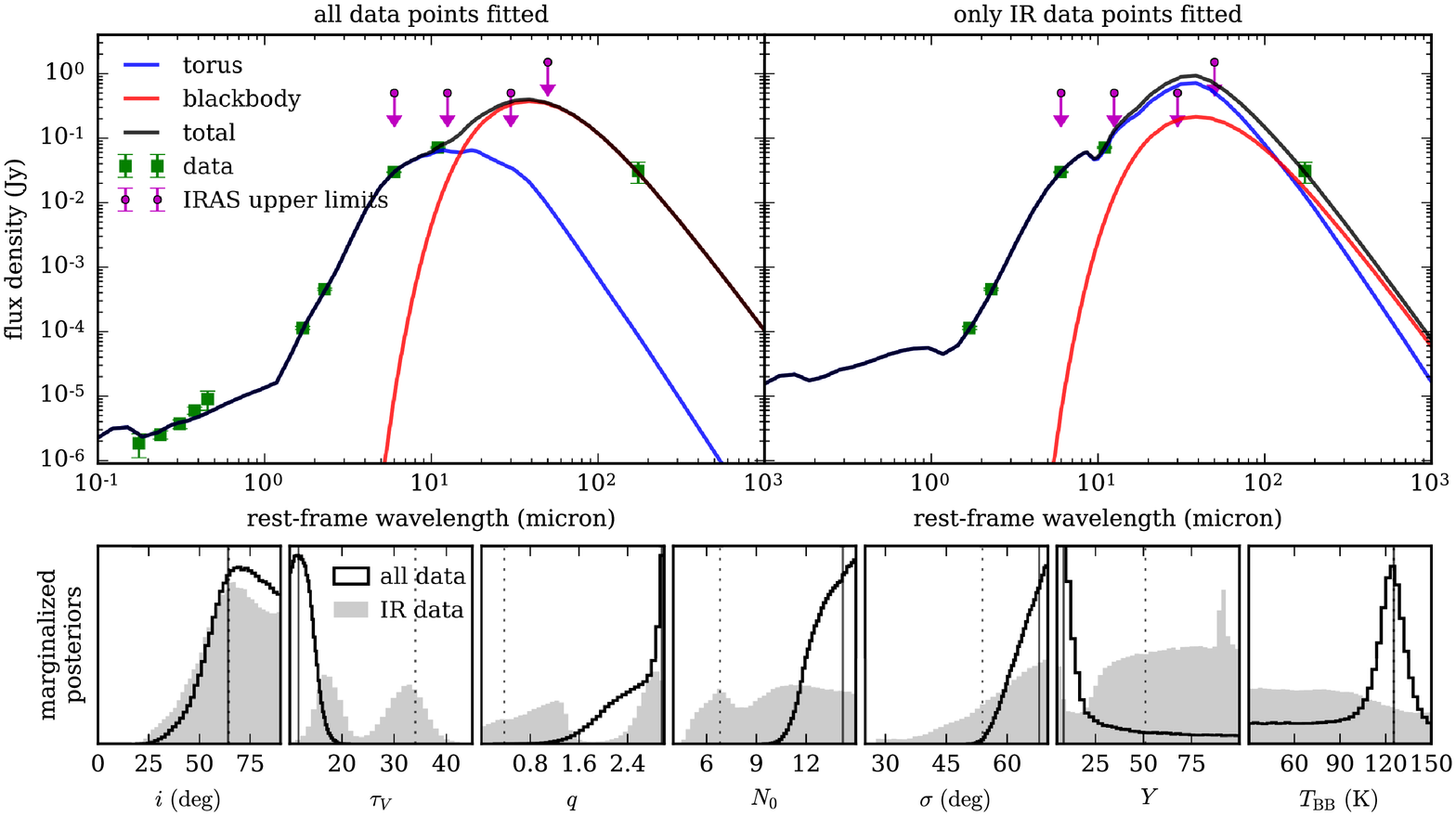} 
  \caption{SED fitting and Bayesian inference of model
    parameters. {\it Upper left:} Maximum-a-posteriori (MAP)
    composite model (in black), comprising an AGN torus component
    (blue) and modified BB (red), fitted to all available data of
    WISE~J1036+0449 (green squares with error bars). The purple circles show the {\it IRAS} upper limits \citep{Neugebauer:1984gf}. {\it Upper
      right:} Same, but fitted to IR data only. {\it Lower panels:}
    Marginalized posteriors of all model parameters, for the all-data
    model (black histogram) and the IR-only model (grey histogram). The solid and
    dotted vertical lines indicate the MAP values of each
    parameter, for the two respective models. The SED is in $F_{\nu}$ for consistency with the fitting procedure of \citet{Nikutta2012phd}. A detailed description of the different parameters and of the results obtained by the SED fitting is reported in Appendix\,\ref{sect:SEDmodel} and \ref{sect:SEDfittResults}.}
  \label{fig:fitting}
\end{figure*}
%

A notable difference in the spectral shapes of the two models is that
in the all-data fit, the torus component produces a 10\mic\ silicate
feature in weak emission, while the MAP model obtained with IR data
alone shows a clear silicate absorption feature at 10\mic. Obtaining a
spectrally resolved SED in that wavelength region could thus greatly
assist in constraining the range of likely models. The posterior distributions of all parameters are shown in the lower
panels of Fig.~\ref{fig:fitting}. The empty histograms are for the
all-data model in the left upper panel, the grey ones for the IR-only
model. Vertical lines indicate the MAP values of each parameter for
the two respective models. It should be remarked that the IR-only
model would predict a flux higher than that of the sources studied by \cite{Tsai:2015qf} in the 15--70\,$\mu$m range. A detailed description of the results obtained by the SED fitting is reported in Appendix\,\ref{sect:SEDfittResults}.

\section{X-ray Observations}\label{Sect:data_analysis}

\subsection{NuSTAR}\label{Sect:nustar_data_analysis}
The {\it Nuclear Spectroscopic Telescope Array} ({\it NuSTAR}, \citealp{Harrison:2013uq}) carried out three observations of WISE\,J1036+0449 between 2014 November 1 and December 22 (see Table\,\ref{tab:obslog}). The data obtained by the two focal plane modules (FPMA and FPMB) were processed using the {\it NuSTAR} Data Analysis Software \textsc{nustardas}\,v1.4.1 within Heasoft\,v6.16, using the calibration files released on UT 2015 March 16 \citep{Madsen:2015uq}. For each observation the cleaned and calibrated event files, together with the exposure maps, were produced using \textsc{nupipeline} following the standard guidelines. For each focal plane module we merged the images of the three observations in the 3--24\,keV band using \textsc{Ximage}, taking into account the exposure maps. The total on-source exposure time is 68.5 and 68.4\,ks for FPMA and FPMB, respectively. The source is detected both in the 3--10\,keV and 10--24\,keV band (at 4.6$\sigma$ and 3$\sigma$, respectively). In the combined 3--24\,keV image (Fig.\,\ref{fig:images}) the source is detected at the $\sim 6\sigma$ level.

The source spectra and light-curves were extracted using the \textsc{nuproducts} task, selecting circular regions of 45\,arcsec radius centred on the position of the source reported in the {\it WISE} catalog (RA, DEC: 159.20133$^{\circ}$, 4.83086$^{\circ}$), while for the background spectra and light-curves we used an annular region centred on the source with inner and outer radii of 90 and 150\,arcsec, respectively. The source and background spectra, together with the \textsc{rmf} and \textsc{arf} files for the three observations, were merged using the \textsc{addascaspec} task for both FPMA and FPMB. We then merged the FPMA and FPMB spectra, background and responses. The final {\it NuSTAR} spectrum of WISE\,J1036+0449 is shown in the left panel of Fig.\,\ref{fig:xrayspec_myt}. The two FPM cameras detected a total of $\sim 120$ background-subtracted counts.

\subsection{Swift/XRT}\label{Sect:xrt_data_analysis}
The X-ray telescope (XRT, \citealp{Burrows:2005vn}) on board {\it Swift} \citep{Gehrels:2004dq} carried out three short (1--2\,ks each) observations of WISE\,J1036+0449, approximately simultaneously with {\it NuSTAR}. {\it Swift}/XRT data analysis was performed using the \textsc{xrtpipeline} following the standard guidelines. We inspected the combined {\it Swift}/XRT images in the 0.3--1.5\,keV, 1.5--10\,keV and the 0.3--10\,keV (Fig.\,\ref{fig:images}) bands, and did not find any evidence of a detection of the source.

\section{X-ray spectral analysis}\label{sect:specAnalysis}
The X-ray spectral analysis was performed using \textsc{xspec}\,v.12.8.2 \citep{Arnaud:1996kx}. We added to all models Galactic absorption in the direction of the source ($N_{\rm\,H}^{\rm\,Gal}=2.9\times 10^{20}\rm\,cm^{-2}$, \citealp{Kalberla:2005fk}), using the \textsc{TBabs} model \citep{Wilms:2000vn}. We set the abundances to solar values. Due to the low signal-to-noise ratio of the observations, we used the Cash statistic \citep{Cash:1979fk} to fit the data. The source spectrum was binned to have two counts per bin, in order to avoid issues related to empty bins in \textsc{xspec}.

As a first test, we fitted the spectrum with a simple power-law model. This resulted in a photon index of $\Gamma = 1.2\pm 0.5$, lower than the value typically found for unobscured AGN (e.g., \citealp{Nandra:1994ly,Piconcelli:2005tg,Ricci:2011oq}), and symptomatic of heavy line-of-sight obscuration.

\begin{figure*}[t!]
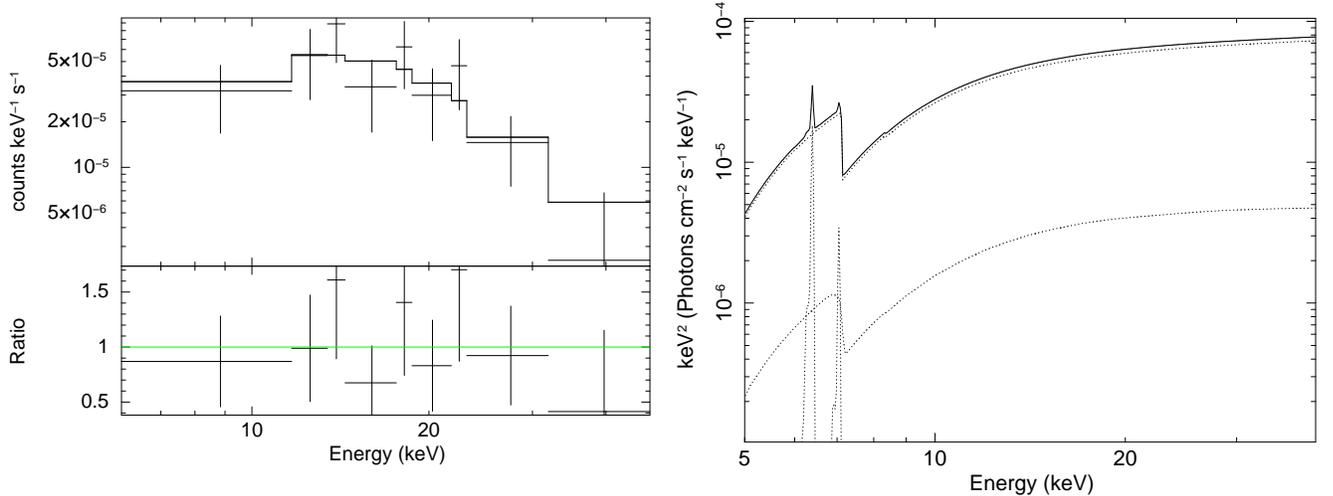

\centering
\begin{minipage}[!b]{.48\textwidth}
\centering
\includegraphics[height=9cm,angle=270]{WISE_myt_ldata.ps}\end{minipage}
\begin{minipage}[!b]{.48\textwidth}
\centering
\includegraphics[height=9cm,angle=270]{myt_model.ps}\end{minipage}
 \begin{minipage}[t]{\textwidth}
  \caption{{\it Left panel:} combined {\it NuSTAR} FPMA and FPMB spectrum of WISE\,J1036+0449 fitted with the \textsc{MYTorus} model (assuming $\Gamma=1.9$, see $\S$\ref{sect:specAnalysis}). The bottom panel shows the ratio between the data and the model. {\it Right panel}: spectrum of the \textsc{MYTorus} model with the parameters fixed to those obtained by fitting the combined {\it NuSTAR} FPMA and FPMB spectrum. The dotted lines represent the primary X-ray emission (upper part of the figure), the scattered component (lower part) and the fluorescent lines. In both plots the energies are in the rest frame of the source.
}
\label{fig:xrayspec_myt}
 \end{minipage}
\end{figure*}

\subsection{Pexrav}
We then fitted the data with a model that consists of an absorbed power-law with a photon index fixed to $\Gamma=1.9$, consistent with the average intrinsic value of AGN (e.g., \citealp{Nandra:1994ly,Piconcelli:2005tg,Ricci:2011oq}), and unabsorbed reprocessed X-ray emission from a slab (\textsc{pexrav} in \textsc{xspec}; \citealp{Magdziarz:1995pi}). In this scenario the outer part of the accretion disk is producing the bulk of the reprocessed radiation. The value of the reflection component was fixed to $R=0.5$, while the inclination angle was set to $\theta_{\rm\,i}=30^{\circ}$. Obscuration was taken into account by combining Compton scattering ($\textsc{cabs}$) and photoelectric absorption ($\textsc{zphabs}$). Due to the low S/N of the spectra the Fe K$\alpha$ line (at $\sim 3.2$\,keV in the observed frame), expected to arise from reprocessing of the primary X-ray radiation in the circumnuclear material (e.g., \citealp{Shu:2010zr,Ricci:2014vn,Gandhi:2015qf}), could not be detected. In \textsc{xspec} the model is:

\smallskip
\noindent\textsc{tbabs$_{\rm\,Gal}$(zphabs$\times$cabs$\times$zpowerlaw + pexrav)}. 
\smallskip

\noindent The fit results in a value of the Cash statistic of $C=173$ for 170 degrees of freedom (DOF). The value of the line-of-sight column density obtained is $N_{\rm\,H}=5.3^{+11.9}_{-4.8}\times 10^{23}\rm\,cm^{-2}$, while the rest-frame 2--10\,keV intrinsic luminosity is $4.1\times 10^{44}\rm\,erg\,s^{-1}$. Assuming larger ($R=1$) or lower ($R=0.1$) values of the reflection component produces consistent values of the column density ($N_{\rm\,H}\leq 2.0\times 10^{24}\rm\,cm^{-2}$ and $N_{\rm\,H}=6.7^{+9.4}_{-5.0}\times 10^{23}\rm\,cm^{-2}$, respectively). 

We also applied the \textsc{plcabs} model \citep{Yaqoob:1997ez}, which reproduces absorption considering an uniform spherical distribution of matter:

\smallskip
\noindent\textsc{tbabs$_{\rm\,Gal}$(plcabs + pexrav)}. 
\smallskip

\noindent This model also produces a good fit to the data $C/\rm DOF=173.4/170$, and the column density obtained is $N_{\rm\,H}\leq 2.41\times 10^{24}\rm\,cm^{-2}$.

\subsection{MYTorus}\label{sec:mytorus}

Next, we applied the \textsc{MYTorus} model \citep{Murphy:2009ly} to reproduce self-consistently absorption and reflection assuming a smooth torus with a half-opening angle of $\theta_{\mathrm{OA}} = 60^{\circ}$. The model is composed of three additive and exponential table models: the zeroth-order continuum (\textsc{MYTorusZ}), the scattered continuum (\textsc{MYTorusS}), and a component containing the fluorescent emission lines (\textsc{MYTorusL}). In \textsc{xspec} the model we used is:

\smallskip
\noindent \textsc{tbabs$_{\rm\,Gal}\times$\{MYTorusZ$\times$ zpowerlaw + MYTorusS + MYTorusL\}}. 
\smallskip

\noindent Applying \textsc{MYTorus}, fixing $\Gamma=1.9$ and the inclination angle to $\theta_{\rm\,i}=90^{\circ}$ (which corresponds to an edge-on view in this geometry), we obtained $C=172.7$ for 170 DOF and a value of the column density consistent with the one inferred using \textsc{pexrav} ($N_{\rm\,H}=7.1^{+8.1}_{-5.1}\times 10^{23}\rm\,cm^{-2}$). The X-ray spectrum of WISE\,J1036+0449 and the model used for the fit (both assuming $\Gamma=1.9$) are shown in Fig.\,\ref{fig:xrayspec_myt} (left and right panel, respectively). The rest-frame 2--10\,keV intrinsic luminosity of the best-fit \textsc{MYTorus} model is $6.3\times 10^{44}\rm\,erg\,s^{-1}$.

\begin{figure*}[t!]
\centering
\begin{minipage}[!b]{.48\textwidth}
\centering
\includegraphics[width=9cm]{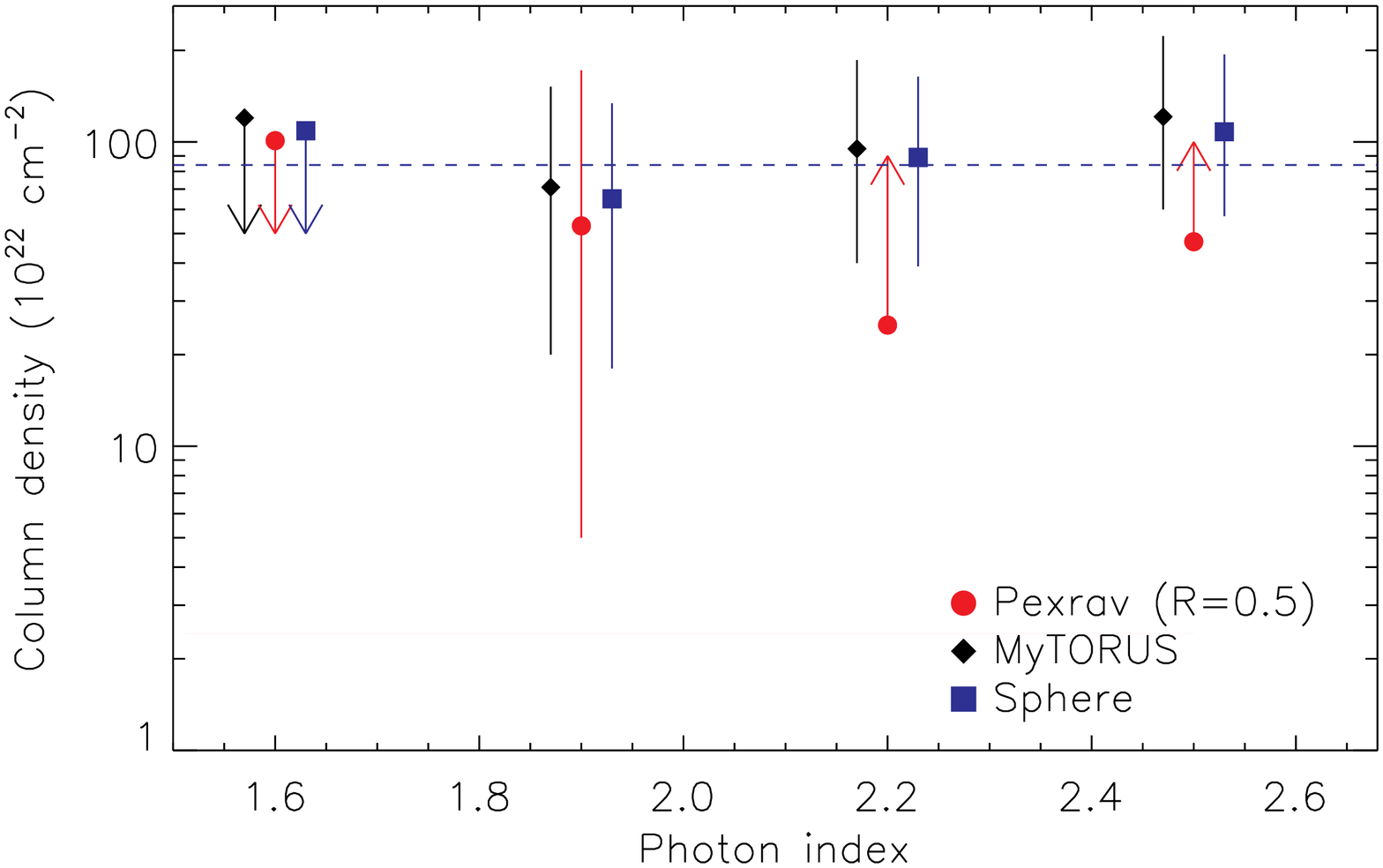}\end{minipage}
\begin{minipage}[!b]{.48\textwidth}
\centering
\includegraphics[height=9cm,angle=90]{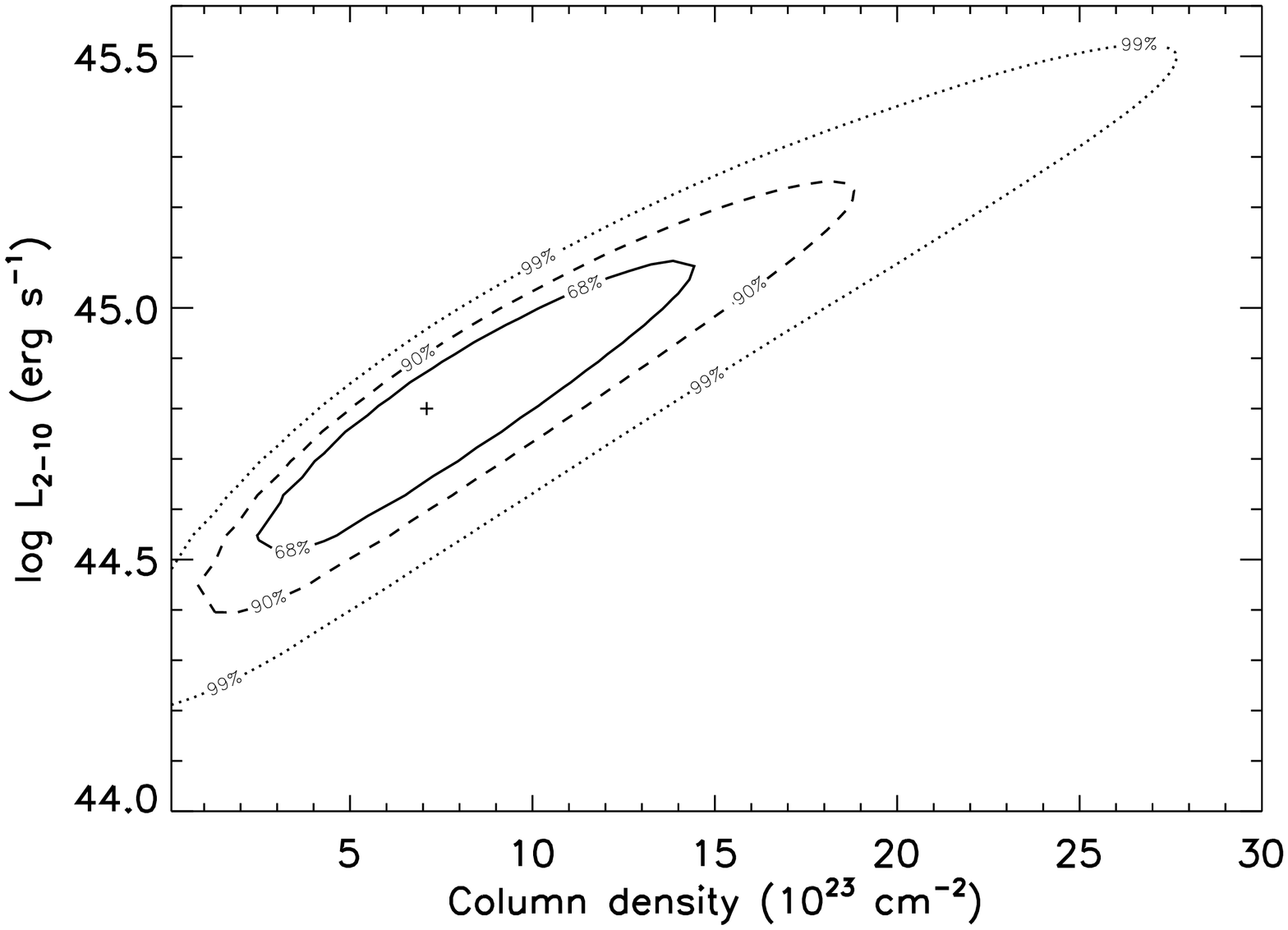}\end{minipage}
 \begin{minipage}[t]{\textwidth}
  \caption{{\it Left panel}: Values of the column density $N_{\rm\,H}$ obtained by assuming different values of the photon index of the primary X-ray radiation for the \textsc{pexrav}, \textsc{MYTorus} and \textsc{sphere} models. The points obtained with the last two models have been shifted by $\Delta \Gamma=\mp0.03$ for visual clarity. The dashed line shows the value of $N_{\rm\,H}$ obtained from $E(B-V)$ by using Eq.\,\ref{eq:NHebv} (see also Table\,\ref{tab:NH_Lx_hotdogs}). {\it Right panel}: Confidence intervals of the intrinsic 2--10\,keV luminosity and the column density for WISE\,J1036+0449 obtained using the \textsc{MYTorus} model. The continuous, dashed and dotted lines represent the 68\%, 90\% and 99\% confidence contours, respectively. The black cross represents the best-fit value of the parameters. }
\label{fig:gammaNH}
\end{minipage}
\end{figure*}

\subsection{Sphere}\label{sect:sphere}

We then applied the \textsc{sphere} model \citep{Brightman:2011oq}, which assumes that the X-ray source is fully covered by the obscuring material. The physical scenario associated with this model is that the AGN is expelling quasi-isotropically the circumnuclear material because of the high radiation pressure, similarly to what was proposed by \citet{Diaz-Santos:2016ly} for WISE\,J2246$-$0526. In \textsc{xspec} the model is: 

\smallskip
\noindent \textsc{tbabs$_{\rm\,Gal}$(atable\{sphere0708.fits\})}.  
\smallskip

\noindent We fixed $\Gamma=1.9$, and also found this model to give a good fitÕ ($C=173.1$ for 170 DOF), supporting that the source is heavily obscured ($N_{\rm\,H}=6.5^{+6.9}_{-4.7}\times 10^{23}\rm\,cm^{-2}$). The rest-frame 2--10\,keV intrinsic luminosity obtained with \textsc{sphere} is $4.0\times 10^{44}\rm\,erg\,s^{-1}$.

\section{Discussion}\label{sect:discussion}
We have reported here on the results obtained from the study of one of the closest Hot DOGs known, WISE\,J1036+0449. In the following we discuss the obscuration ($\S$\ref{sect:obscuration}) and X-ray ($\S$\ref{sect:xrweakness}) properties of this object, and of those of Hot DOGs in general.

\subsection{Obscuration in Hot DOGs}\label{sect:obscuration}
Studying the luminosity function of Hot DOGs, \cite{Assef:2015zr} have shown that Hot DOGs have a similar space density as the most luminous ($L_{\rm\,Bol}\gtrsim 10^{47}\rm\,erg\,s^{-1}$) unobscured quasars. It is therefore important to constrain the absorption properties of this significant population of obscured quasars. We have shown in $\S\ref{sect:selection}$ that WISE\,J1036+0449 and Hot DOGs at higher redshifts have very similar multi-wavelength characteristics, and they belong to the same class of AGN.

For the three X-ray spectral models discussed in $\S$\,\ref{sect:specAnalysis} we also tested values of $\Gamma=1.6$, $\Gamma=2.2$ and $\Gamma=2.5$, finding that, depending on the shape of the X-ray continuum, the value of $N_{\rm\,H}$ varies between $N_{\rm\,H}\leq 1.2\times 10^{24}\rm\,cm^{-2}$ and $N_{\rm\,H}\geq 4.7\times 10^{23}\rm\,cm^{-2}$ (Fig.\,\ref{fig:gammaNH}). Steeper slopes imply higher values of the column density. As shown in the figure (see also Table\,\ref{tab:NH_Lx_hotdogs}), this range of $N_{\rm\,H}$ is in agreement with the column density estimated from the extinction $E(B-V)$, assuming the relation
\begin{equation}\label{eq:NHebv}
\frac{E(B-V)}{N_{\rm\,H}}=1.5\times10^{-23}\rm\,cm^2\,mag
\end{equation}
reported by \citeauthor{Maiolino:2001zr} (\citeyear{Maiolino:2001zr}; see also \citealp{Burtscher:2016dq} for a recent discussion on the subject).

A well-known correlation exists between the photon index and the Eddington ratio ($\lambda_{\rm\,Edd}$), with the slope varying depending on the range of $\lambda_{\rm\,Edd}$ probed (see Fig.\,1 of \citealp{Zhou:2015uq}). For values of $\lambda_{\rm\,Edd}$ in the range $10^{-2.6}-1$ the correlation is positive (e.g., \citealp{Shemmer:2006zr,Brightman:2013ys,Brightman:2016oj}). For $\log \lambda_{\rm\,Edd} \geq -0.7$ the relation between $\Gamma$ and $\lambda_{\rm\,Edd}$ seems to be different than at lower values of $\lambda_{\rm\,Edd}$, and the average photon index is $\Gamma\sim 1.9$ (e.g., \citealp{Ai:2011kx,Kamizasa:2012fk,Ho:2016ys}). \cite{Assef:2015zr} have shown that, unless Hot DOGs deviate significantly from the local $M-\sigma$ relation, they radiate above the Eddington limit, with typical values of $\lambda_{\rm\,Edd}\gtrsim 2$ (see Fig.\,8 of \citealp{Assef:2015zr} and \citealp{Tsai:2015qf} for discussions on the subject). The bolometric luminosity of WISE\,J1036+0449 from the SED is $L_{\rm\,Bol}\simeq 8\times 10^{46}\rm\,erg\,s^{-1}$, and considering the black hole mass estimated from the broadened Mg\,{\sc ii} ($M_{\rm\,BH}\simeq 2 \times 10^8 M_{\odot}$) we find that the source is accreting above the Eddington limit ($\lambda_{\rm\,Edd}\simeq 2.7$). It should be stressed that even an outflow origin of the broadening of Mg\,{\sc ii} would imply that the source is accreting at high values of the Eddington ratio. Therefore, we expect that $\Gamma\sim 1.9$ is a reasonable assumption for WISE\,J1036+0449 and for Hot DOGs in general.

Although, by construction, the selection function of Hot DOGs would not identify low-obscuration objects, it should be able to select objects with $N_{\rm\,H}> 10^{24}\rm\,cm^{-2}$.  As discussed by \cite{Assef:2015zr}, considering the values of $E(B-V)$ obtained by their analysis and Eq.\,\ref{eq:NHebv}, the typical column densities of Hot DOGs are expected to be in the range $1.7\times 10^{23}<N_{\rm\,H}<1.4\times 10^{24}\rm\,cm^{-2}$. This is very different from the intrinsic column density distribution of local, less-luminous AGN, with the $N_{\rm\,H}$ distribution of the former showing a sharp peak in the $\log (N_{\rm\,H}/\rm cm^{-2})=23-24$ range and very few CT objects, while local AGN have a significantly more uniform distribution, and a fraction of $27\pm 4\%$ of Compton-thick (CT, $N_{\rm\,H}\geq 10^{24}\rm\,cm^{-2}$) AGN (\citealp{Ricci:2015fk}, see also \citealp{Koss:2016os}). It should be remarked, however, that \cite{Ricci:2015fk} have also shown that the fraction of local hard X-ray selected CT AGN decreases with increasing values of the luminosity. To date, only a few direct measurements of the line-of-sight column density of Hot DOGs have been performed (see Table\,\ref{tab:NH_Lx_hotdogs}), and they seem to be consistent with what was inferred by \cite{Assef:2015zr}, albeit with large uncertainties. The hyperluminous quasar IRAS\,09104+4109 ($L_{\rm\,IR}\simeq 5.5\times 10^{46}\rm\,erg\,s^{-1}$), recently observed by {\it NuSTAR}, also shows a column density in a similar range ($N_{\rm\,H}\sim5\times10^{23}\rm\,cm^{-2}$, \citealp{Farrah:2016kq}), while the HyLIRG IRAS\,F15307+3252 is significantly more obscured [$N_{\rm\,H}\gtrsim 2.5\times10^{24}\rm\,cm^{-2}$, \cite{Hlavacek-Larrondo:2016lq}] and shows a strong Fe\,K$\alpha$ line ($EW\sim 1-3$\,keV).

A possible explanation for the difference between the column density distribution of local AGN and that of Hot DOGs obtained from $E(B-V)$ could be related to differences in the dust-to-gas ratios. The circumnuclear material in Hot DOGs might in fact be significantly more gaseous due to the higher luminosity, which would cause most of the dust in the inner few parsecs to sublimate. Fitting the X-ray spectrum of WISE\,J1036+0449 with the \textsc{sphere} model, setting $N_{\rm\,H}=10^{25}\rm\,cm^{-2}$, we found a worse fit ($C=194.7$ for 171 DOF) than that reported in $\S$\ref{sect:sphere}. It should be remarked that \cite{Gandhi:2016kq} have recently shown, studying the broad-band X-ray spectrum of the LIRG NGC\,7674, that even objects with weak Fe\,K$\alpha$ lines could be heavily obscured and reflection-dominated. The model \cite{Gandhi:2016kq} used for NGC\,7674 can well reproduce the spectrum of WISE\,J1036+0449 ($C=175.0$ for 171 DOF). It might therefore be possible that WISE\,J1036+0449 and other Hot DOGs are significantly more obscured than inferred by current X-ray spectral analysis.

\begin{table*}
\begin{center}
\caption[]{X-ray properties of Hot DOGs.}
\label{tab:NH_Lx_hotdogs}
\begin{tabular}{cccccccccc}
\noalign{\smallskip}
\hline \hline \noalign{\smallskip}
(1) & \multicolumn{1}{c}{(2)} & \,\,\,\,\,(3) & (4) & (5) & (6)  & (7)  & (8) & (9) \\
\noalign{\smallskip}
Source & Redshift & Facility &$N_{\rm\,H}$ &$N_{\rm\,H}^{\rm\,Ext.}$ & $E(B-V)$ &$\log L_{2-10}$  & $L_{6\rm\,\mu m}$ & Reference \\
 &   &   & ({\scriptsize $10^{23}\rm\,cm^{-2}$}) & ({\scriptsize $10^{23}\rm\,cm^{-2}$}) & ({\scriptsize mag})  & ({\scriptsize $\rm\,erg\,s^{-1}$}) & ({\scriptsize $\rm\,erg\,s^{-1}$}) &  \\
\noalign{\smallskip}
\hline \noalign{\smallskip}
W0204$-$0506$^{\rm a}$ &  2.100 &  C		&  $6.3^{+8.1}_{-2.1}$ & $6.5\pm 0.8$	& $9.7\pm1.2$	&	44.90 [44.78 -- 45.34] & 46.86	& \citet{Assef:2016qf}	\\  
\noalign{\smallskip}
W1036+0449$^{\rm a}$ & 1.009 & N	 &  $7.1^{+8.1}_{-5.1}$ & $8.4\pm 0.3$	&  $12.6\pm0.4$	  &	44.80 [44.52 -- 45.09]	 & 46.61   & This Work	\\  
\noalign{\smallskip}
W1814+3412$^{\rm b}$ & 2.452 &   X	& \nodata  & $10.1\pm 1.0$	&   $15.1\pm1.1$ 	& 44.84 [44.61 -- 44.98]  & 47.30 &  \citet{Stern:2014kx}	\\  
\noalign{\smallskip}
W1835+4355$^{\rm a}$ &  2.298 &  X		& $\gg 10$ & $2.9\pm 0.2$	& $4.4\pm 0.3$	& 44.85	 &	46.95 & Zappacosta et al. (in prep.)		\\  
\noalign{\smallskip}
W2207+1939$^{\rm b}$ & 2.021 &  X		& \nodata  & $11.7\pm1.5$	&  $17.6\pm2.3$	 & $\leq 44.78$	   &	 46.92 &  \citet{Stern:2014kx}	\\  
\noalign{\smallskip}
W2357+0328$^{\rm b}$ & 2.113 & X	& \nodata  &	$3.7\pm0.3$ &  $5.5\pm0.4$ 	& 44.52 [44.20 -- 44.65]   &	46.70 &  \citet{Stern:2014kx}	\\  
\noalign{\smallskip}
\hline
\noalign{\smallskip}
\end{tabular}
\tablecomments{The table reports (1) the list of Hot DOGs that have been observed in the X-ray band, (2) the values of the spectroscopic redshift, (3) the facility used (C: {\it Chandra}; N: {\it NuSTAR}; X: {\it XMM-Newton}), (4) the column density inferred from the X-ray spectral analysis, (5) the column density obtained from $E(B-V)$ using Eq.\,\ref{eq:NHebv}, (6) the extinction, (7) the intrinsic rest-frame 2--10\,keV luminosity and the 68\% confidence interval, (8) the rest-frame 6\,$\mu$m luminosity and (9) the reference. The objects reported in the table are typically at lower redshifts and have lower luminosities than those reported in \citet{Tsai:2015qf}\newline
$^{\rm a}$ Sources for which the column density and the intrinsic 2--10\,keV luminosity were inferred from the X-ray spectral analysis. For WISE\,J1036+0449 we reported the value obtained using the \textsc{MYTorus} model (see $\S$\ref{sec:mytorus}). \newline
$^{\rm b}$ Sources for which the column density was obtained from $E(B-V)$ using Eq.\,\ref{eq:NHebv}, and the intrinsic 2--10\,keV rest frame luminosity was extrapolated as illustrated in Appendix\,\ref{sect:appendix2}.
}
\end{center}
\end{table*}

 Another explanation for the different column density distributions could be the following. For a given mass of gas and dust, and assuming a homogeneous distribution of the material, it is more difficult to have Compton-thick [$\log (N_{\rm\,H}/\rm cm^{-2})\geq 24$] lines-of-sights if the inner radius of the absorbing material is at larger distances from the X-ray source than for local AGN. The inner radius of the dust around Hot DOGs can be calculated from the dust sublimation radius (e.g., \citealp{Nenkova:2008cr}):
\begin{equation}\label{eq:dustradius}
R_{\rm\,d}\simeq 0.4\left(\frac{L_{\rm\,bol}}{10^{45}\rm\,erg\,s^{-1}}\right)^{0.5}\left(\frac{1500\rm\,K}{T_{\rm\,sub}}\right)^{2.6}\rm\,pc,
\end{equation}
where $T_{\rm\,sub}$ is the sublimation temperature of dust grains. Considering silicate dust grains ($T_{\rm\,sub}=1500$K), and the average bolometric luminosity of Hot DOGs reported by \cite{Assef:2015zr} ($\langle L_{\rm\,bol}\rangle \simeq 5\times10^{47}\rm\,erg\,s^{-1}$), we obtain that the inner radius of the dust is $\sim 9\rm\,pc$.  This is much larger than the typical value expected for {\it Swift}/BAT AGN ($R_{\rm\,d}\sim0.1-0.4\rm\,pc$ considering $L_{\rm\,Bol}\sim 10^{44}-10^{45}\rm\,erg\,s^{-1}$). It should be remarked that, while these distances might be systematically smaller by a factor of $\sim 3$, as found by near-IR reverberation studies \citep{Kishimoto:2007cr}, the ratio between $R_{\rm\,d}$ of local AGN and Hot DOGs would be the same. Therefore, if Hot DOGs have dust masses comparable to local AGN, and gas and dust are strongly coupled, a difference in the column density distribution would be expected. While there might be a significant fraction of dust-free material within the sublimation radius, this material is also likely to be highly photo-ionized by the very luminous AGN. Moreover, as shown by \citet{Diaz-Santos:2016ly} for WISE\,J2246$-$0526, it is possible that in Hot DOGs, due to the strong radiation pressure, the gas is being blown away isotropically.

Hot DOGs could be very different from local, less-luminous AGN, where the bulk of the gas and dust is believed to be distributed in a torus-like structure, and might represent a short-lived transition phase between heavily obscured and unobscured AGN \citep{Bridge:2013qa}, similar to red quasars (e.g., \citealp{Urrutia:2008vn,Banerji:2012mz,LaMassa:2016ly}). The idea that this phase might happen following a major merger is supported by the recent results of \citep{Fan:2016sf} using {\it Hubble Space Telescope}/WFC3 images. Studying SCUBA observations \cite{Jones:2015lq} showed that Hot DOGs have an excess of SMG neighbours, which would increase the chances of a merger. These results are in agreement with the idea that the AGN unification model might not be valid at high luminosities, where galaxy mergers are more important than secular processes in triggering accretion onto the SMBH (e.g., \citealp{Treister:2012fk}).

\bigskip
\bigskip

\subsection{Are Hot DOGs intrinsically X-ray weak?}\label{sect:xrweakness}

AGN show a strong positive correlation between \mbox{2--10\,keV} and mid-IR luminosity (at 6\,$\mu$m or 12\,$\mu$m, e.g. \citealp{Lutz:2004vn,Ichikawa:2012fk,Mateos:2015tg}), as confirmed by high angular resolution ($\sim 0.35$\,arcsec) mid-IR studies (e.g., \citealp{Gandhi:2009dq}, \citealp{Levenson:2009kl}, \citealp{Asmus:2015ly}) of AGN in the {\it Seyfert regime} ($L_{6\,\mu\rm m}<10^{44}\rm\,erg\,s^{-1}$). In the {\it quasar regime} ($L_{6\,\mu\rm m}>10^{44}\rm\,erg\,s^{-1}$),  \cite{Stern:2015vn} found evidence of a flattening of the relationship for $L_{6\,\mu\rm m} \gtrsim 10^{46}\rm\,erg\,s^{-1}$, with most sources being fainter than expected in the X-ray band, in agreement with what was found by \cite{Fiore:2009hc} and \cite{Lanzuisi:2009ij}. To reproduce this trend, \cite{Stern:2015vn} proposed a revised formulation, using a second-order polynomial to fit the data.

A deviation of the mid-IR/X-ray correlation at high luminosities might be expected, considering the following arguments. The flux in the mid-IR is believed to be due to reprocessing of optical, UV and Extreme UV photons by the gas and dust in the putative molecular torus; therefore the main driver of the mid-IR/X-ray correlation is the relation between the optical/UV and X-ray flux, which has been analyzed by studies focussed on two related quantities: i) the optical to X-ray flux ratio ($\alpha_{\rm\,OX}$), which is the ratio between the monochromatic 2\,keV and 2500\AA\ luminosities; ii) the 2--10\,keV bolometric correction ($\kappa_{\mathrm{x}}$). Recent works have shown that both $\alpha_{\rm\,OX}$ (e.g., \citealp{Lusso:2010bs}) and $\kappa_{\mathrm{x}}$ (e.g., \citealp{Vasudevan:2007fv}) depend on $\lambda_{\rm\,Edd}$, with the optical flux increasing with respect to the X-ray flux for higher values of $\lambda_{\rm\,Edd}$. This effect could be related to the different physics of the accretion flow and corona, or to the fact that the X-ray source is saturated by the high rate of optical/UV photons produced by the accretion flow.

According to the relation of \cite{Stern:2015vn}, the expected 2--10\,keV luminosity of WISE\,J1036+0449 would be $\log (L_{2-10\rm\,keV}/\rm erg\,s^{-1})\simeq 45.25$, a value $\sim 3$ times higher than that obtained by our X-ray spectral analysis [$\log (L_{2-10\rm\,keV}/\rm erg\,s^{-1})\simeq 44.80$]. To constrain the relation between the 6$\,\mu$m and 2--10\,keV luminosity for the three objects from \citet{Stern:2014kx} that have been observed so far by X-ray facilities, but for which no spectral analysis could be performed because of the low significance of the detections, we calculated the intrinsic 2--10\,keV luminosity from the value of $N_{\rm\,H}$ obtained from $E(B-V)$ and the 0.5--10\,keV observed flux. The procedure adopted is described in Appendix\,\ref{sect:appendix2}. The values of $L_{2-10\rm\,keV}$, $L_{6\,\mu\rm m}$, $N_{\rm\,H}$ and $E(B-V)$ for all Hot DOGs observed in the X-rays are listed in Table\,\ref{tab:NH_Lx_hotdogs}. A possible caveat of this approach is that, as mentioned in $\S$\ref{sect:obscuration}, it is still largely unknown whether in the extreme environments of Hot DOGs the relationship between $E(B-V)$ and $N_{\rm\,H}$ is consistent with that found for local AGN by \cite{Maiolino:2001zr}. However, it should be remarked that the values of the column density obtained for WISE\,J0204$-$0506 and WISE\,J1036+0449 through X-ray spectral analysis are in agreement with those extrapolated from $E(B-V)$ (see Table\,\ref{tab:NH_Lx_hotdogs} and Fig.\,\ref{fig:gammaNH}). A possible exception is WISE\,J1835+4355, which is significantly more obscured than what would be predicted by $E(B-V)$ (Zappacosta et al. in prep.).

As shown in Fig.\,\ref{fig:lxvslmir_stern}, the absorption-corrected X-ray luminosities of all the Hot DOGs (red circles) observed by X-ray facilities are significantly lower than the values expected by the mid-IR/X-ray correlation, which implies that they might be either intrinsically X-ray weak or significantly more obscured than estimated. Hot DOGs are also significantly less luminous in the X-ray band than the unobscured quasars with similar 6\,$\mu$m luminosities shown in the figure (from \citealp{Just:2007qf}). This is found in objects for which $L_{2-10\rm\,keV}$ was obtained from spectral analysis, as well as in those for which we used the indirect approach described in Appendix\,\ref{sect:appendix2}. Considering the average bolometric luminosity of Hot DOGs \citep{Assef:2015zr}, $\langle L_{\rm\,bol}\rangle \simeq 5\times10^{47}\rm\,erg\,s^{-1}$, the Eddington ratio would be $\lambda_{\rm\,Edd}\simeq 1$ even assuming an average black hole mass of $M_{\rm\,BH}\sim 4\times 10^9\,M_{\odot}$. If so, $\kappa_{\mathrm{x}}$ would be $\simeq100$, a value 5--10\,times larger than for $\lambda_{\rm\,Edd}\lesssim 0.1$ \citep{Vasudevan:2009dz}. For WISE\,J1036+0449 we found that $\lambda_{\rm\,Edd}\simeq 2.7$ and, considering the bolometric luminosity inferred from the SED ($L_{\rm\,Bol}\simeq 8\times 10^{46}\rm\,erg\,s^{-1}$), we find that $\kappa_{\mathrm{x}}\simeq 130$. It should be remarked that $L_{\rm\,Bol}$ was calculated by interpolating with a power-law the {\it WISE} and CSO data, which might underestimate the real value of the bolometric output for this source and therefore the value of $\kappa_{\mathrm{x}}$. The variation of $\kappa_{\mathrm{x}}$ and $\alpha_{\rm\,OX}$ could therefore straightforwardly lead to the observed deviation in the mid-IR/X-ray correlation at high values of $\lambda_{\rm\,Edd}$.

X-ray weakness has been found to be rather common in broad-absorption line quasars (e.g., \citealp{Gallagher:2001pi,Luo:2013fk,Luo:2014kl}), and, recently, it has been discussed that it might be found also in some ULIRGs (e.g., \citealp{Teng:2014kl,Teng:2015oq}). However, the mechanism responsible for the quenching of the X-ray emission is still unknown. A possible explanation is that the black-hole masses of Hot DOGs are smaller than those of unobscured quasars with similar luminosities, which would lead to larger values of $\lambda_{\rm\,Edd}$, and therefore of $\kappa_{\mathrm{x}}$ and $\alpha_{\rm\,OX}$. 
\cite{Luo:2013fk,Luo:2014kl} argued that the X-ray weakness of broad-absorption line quasars would substitute the shielding material often invoked to prevent the overionization of the wind from the X-ray radiation, thus leading to the launching of more powerful winds. Alternatively, it has been proposed by \citet{Proga:2005ff} that outflows from the accretion disk could collide with the corona, suppressing the production of X-ray emission. Considering the extreme luminosities and Eddington ratios of Hot DOGs, this mechanism appears plausible, also in light of the recent ALMA study of \cite{Diaz-Santos:2016ly}, who found evidence of extremely powerful winds in the most luminous Hot DOG known.

\begin{figure}
  \begin{center}
    \plotone{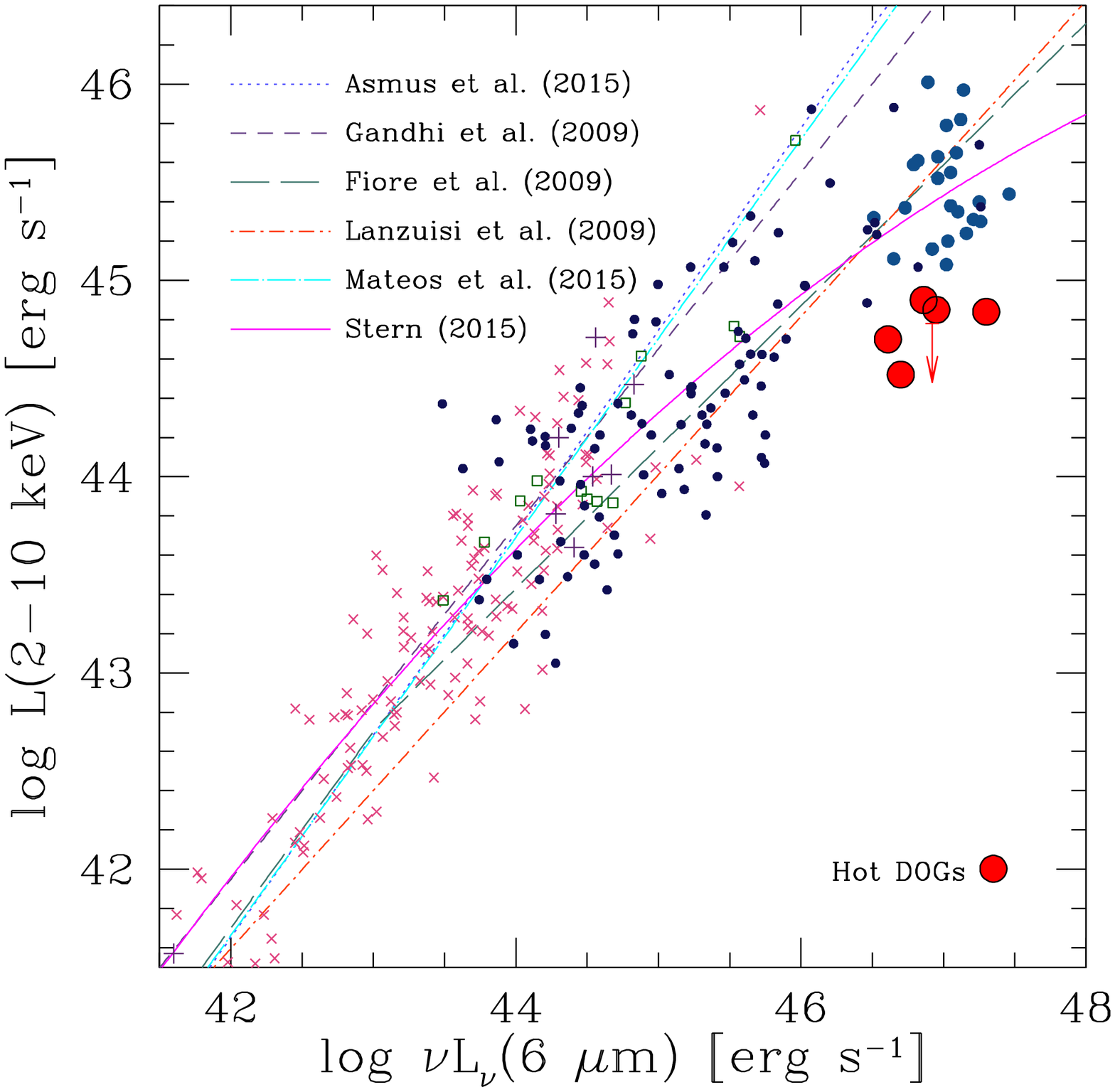}
    \caption{Rest-frame 6\,$\mu$m luminosity versus intrinsic rest-frame 2--10\,keV luminosity for several samples of AGN: Hot DOGs from this work (big red circles, the red arrow represents the upper limit for WISE\,J2207+1939), broad-lined AGN from the SEXSI survey (\citealp{Eckart:2010ve}; open green squares), luminous quasars from \citet{Just:2007qf} (large solid blue circles), Compton-thin AGN observed by {\it NuSTAR} (\citealp{Alexander:2013bh}, \citealp{Del-Moro:2014dq}: purple crosses), quasars from SDSS DR5 (\citealp{Young:2009cr}; small blue dots) and local Seyfert galaxies (\citealp{Asmus:2015ly}, \citealp{Gandhi:2009dq}; red exes). The lines illustrate five $L_{6\,\mu\rm m}$---$L_{2-10\rm\,keV}$ relations (see $\S$\ref{sect:xrweakness} for details). The mid-IR luminosities of \cite{Asmus:2015ly} were obtained at 12\,$\mu$m. The figure shows that Hot DOGs are significantly weaker in the X-ray band than unobscured quasars in the same $6\,\mu$m luminosity range.}
    \label{fig:lxvslmir_stern}
  \end{center}
\end{figure}

\section{Summary and conclusions}\label{sect:conclusion}

We reported here on the multi-wavelength study of WISE\,J1036+0449, the first Hot DOG detected by {\it NuSTAR}. The source was selected using new selection criteria that identify Hot DOGs at lower redshifts than previously discovered. We report below the main findings of our work.

\begin{itemize}

\item The redshift of WISE\,J1036+0449 is $z=1.009$.  The SED of the source is extremely similar to those of Hot DOGs at $z\sim 2$ (Fig.\,\ref{fig:SED}), validating the new method to select Hot DOGs at $z\simeq 1$. 

\item The source is detected in the X-ray band, which confirms the presence of a powerful AGN. We found that the source is obscured [$N_{\rm\,H}\simeq(2-15)\times10^{23}\rm\,cm^{-2}$], with a column density consistent with that of the bulk of the Hot DOG population.

\item If the broadening of the Mg\,{\sc ii} line is due to the gravitational field of the SMBH, then the black hole mass is $M_{\rm\,BH}\simeq 2 \times 10^8 M_{\odot}$ and the Eddington ratio $\lambda_{\rm\,Edd}\simeq 2.7$.

\item The intrinsic 2--10\,keV luminosity of WISE\,J1036+0449 [$\log (L_{2-10\rm\,keV}/\rm erg\,s^{-1})\sim44.80$] is considerably lower than the value expected from the mid-IR/X-ray luminosity correlation, considering its $6\,\mu$m luminosity [$\log (L_{6\rm\,\mu m}/\rm erg\,s^{-1})\sim46.61$], and the 2--10\,keV bolometric correction is $\kappa_{\mathrm{x}}\simeq 130$. Other Hot DOGs are fainter than expected in the X-ray band (Fig.\,\ref{fig:lxvslmir_stern}), which might imply that X-ray weakness is a common characteristic of extremely luminous AGN. X-ray weakness might either be related to significantly larger values of $\lambda_{\rm\,Edd}$ (and therefore of $\kappa_{\mathrm{x}}$ and $\alpha_{\rm\,OX}$), and/or to the disruption of the X-ray corona caused by outflowing material. An alternative explanation is that Hot DOGs are significantly more obscured than what is inferred by current studies based on X-ray spectroscopy and on the analysis of the SED.

\end{itemize}

Future X-ray observations of Hot DOGs at $z\lesssim 1$ will be extremely important to understand whether these objects are intrinsically X-ray weak and to shed light on the conditions of the X-ray emitting plasma around SMBHs at the highest luminosities and accretion rates.

\acknowledgments
We thank the referee for the very prompt report that helped to improve the article.
This work makes use of data products from the {\it Wide-field Infrared Survey Explorer}, which is a joint project of the University of California, Los Angeles, and the Jet Propulsion Laboratory/California Institute of Technology. {\it WISE} is funded by the National Aeronautics and Space Administration. Some of the data presented herein were obtained at the W.M. Keck Observatory, which is operated as a scientific partnership among the California Institute of Technology, the University of California and the National Aeronautics and Space Administration. The Observatory was made possible by the generous financial support of the W.M. Keck Foundation. This work made use of data from the {\it NuSTAR} mission, a project led by the California Institute of Technology, managed by the Jet Propulsion Laboratory, and funded by the National Aeronautics and Space Administration. We thank the {\it NuSTAR} Operations, Software and Calibration teams for support with the execution and analysis of these observations.  This research has made use of the {\it NuSTAR} Data Analysis Software (\textsc{NuSTARDAS}) jointly developed by the ASI Science Data Center (ASDC, Italy) and the California Institute of Technology (Caltech, USA). This work made use of the NASA/ IPAC Infrared Science Archive and NASA/IPAC Extragalactic Database (NED), which are operated by the Jet Propulsion Laboratory, California Institute of Technology, under contract with the National Aeronautics and Space Administration. This material is based upon work supported by the National Aeronautics and Space Administration under Proposal No. 13-ADAP13-0092 issued through the Astrophysics Data Analysis Program. We thank M. Karouzos and C. S. Chang for their comments on the manuscript. We acknowledge financial support from the CONICYT-Chile grants ``EMBIGGEN" Anillo ACT1101 (CR, ET, FEB), FONDECYT 1141218 (CR, FEB), 3140436 (RN) and 1151408 (RA), Basal-CATA PFB--06/2007 (CR, FEB, ET), and the Ministry of Economy, Development, and Tourism's Millennium Science Initiative through grant IC120009, awarded to The Millennium Institute of Astrophysics, MAS (FEB). CR acknowledges support from the China-CONICYT fund. PG thanks the STFC for support [grant reference ST/J003697/2], WNB acknowledges financial support from {\it NuSTAR} subcontract 44A--1092750. PB is supported by a STFC studentship. A.C.  acknowledges support from the ASI/INAF grant I/037/12/0011/13 and the Caltech Kingsley visitor program. T.D-S. acknowledges support from ALMA-CONICYT project 31130005 and FONDECYT\,1151239. SML is supported by an appointment to the NASA Postdoctoral Program at the NASA Goddard Space Flight Center, administered by Universities Space Research Association under contract with NASA.

{\it Facilities:} \facility{Chandra}, \facility{NuSTAR}, \facility{Swift}, \facility{XMM-Newton}, \facility{WISE}, \facility{Keck}.

\bibliographystyle{apj} 
\bibliography{WISE_nustar.bib}

\appendix

\section{Clumpy torus model}\label{sect:SEDmodel}

In the \C\ model \citep{Nenkova+2002,Nenkova+2008a,Nenkova+2008b} the clouds are arranged
around the central illuminating source in an axially symmetric
configuration, and exist across radial distances ranging from the dust
sublimation radius $R_d$ of the constituent dust grains (set by the
source luminosity), to an outer radius $Y \cdot R_{\rm\,d}$, with $Y$ a free
parameter. The local cloud number density (per unit length) varies
with radial and angular coordinates, and is specified by $N_0$, the
mean number of clouds along a radial ray in the equatorial plane. In
the radial direction it declines as $r^{-q}$, with $q$ a free
parameter. In angular direction (equatorial plane to system axis) the
cloud number per line of sight varies as a Gaussian of width
$\sigma$. Finally, the observer's viewing angle $i$, measured from the
torus axis, is the only external model parameter. The modified BB
component, with an emissivity exponent $\beta=1.5$, is often used to
parameterize star-formation contribution at far-IR (FIR)
wavelengths. The parameter $\beta$ has a typical value in the interval $1 < \beta < 2$ \citep{Huang:2014kq}. The only free
parameter of this component is the dust temperature $T_{\rm\,BB}$.

\section{SED fitting approach}\label{sect:SEDfittApproach}

We employed a Bayesian approach for the fitting. {\it Bayes' Theorem}, here in a simplified notation,

\begin{equation}
  \label{eq:bayestheore}
  {\rm Posterior} \equiv p(\vec\theta|\vec D) \propto p(\vec\theta)\, p(\vec D|\vec\theta) \equiv {\rm Prior} \times {\rm Likelihood},
\end{equation}

provides a straight-forward prescription to compute the sought-after {\it posterior} probability density function $p(\vec\theta|\vec D)$ of model parameter values $\vec\theta$, given the observed data vector $\vec D$ (here, the observed SED flux densities). This multi-dimensional posterior is proportional to the product of a {\it prior} PDF $p(\vec\theta)$ of the parameter values (before seeing the data), and the {\it likelihood} $p(\vec D|\vec\theta)$ that the given parameter values generate a model that is compatible with the data. If the uncertainties on $\vec D$ are Gaussian, then $p(\vec D|\vec\theta) \propto \exp(-\chi^2/2)$ \citep[see e.g.][]{Trotta2008}.

We used a Markov-chain Monte Carlo (MCMC) scheme to sample efficiently the 7-dimensional parameter volume. The code was first developed in \citet{Nikutta2012phd} and since then heavily expanded. At each sampling step a torus model SED is generated through multi-dimensional interpolation of the publicly available \C\ hypercube, while the BB
SED is generated on the fly, given the randomly sampled BB temperature. We applied uniform prior PDFs for all model parameters, i.e. $p(\theta_i) = (\Delta\theta_i)^{-1}$, where $\Delta\theta_i$ is the range of parameter values spanned by any single parameter $\theta_i$. The sampling chains are guaranteed to eventually converge toward the target distribution $p(\vec\theta|\vec D)$ \citep{Metropolis+1953,Hastings_1970}. Finally, integration of the multi-dimensional posterior PDF over all but one of the parameters in
$\vec\theta$, yields so-called {\it marginalized posteriors} in 1-d. In \citet[][Appendix therein]{Nikutta2012phd} it is also shown that given the observed SED and the spectral shapes of all model SED components, the relative normalizations of the components are not free parameters, and can in fact be computed analytically. This is the approach employed by the code. The confidence contours obtained by the fit are shown in Fig.\,\ref{fig:contours_clumpy}.

\section{Results of the SED fitting}\label{sect:SEDfittResults}

The MAP values, and posterior medians with
$1\sigma$ confidence intervals are listed in Table\,\ref{tab:fitting}. In both the all-data and IR-only cases, the prevalent viewing angles
$i$ of the torus (measured in degrees from the torus axis) are $\sim 65$ degrees and are compatible with the MAP values. The
posterior distribution medians are also of similar value, and with
comparable $1\sigma$ confidence intervals around them. Most other
parameters react more strongly to the presence or lack of optical/UV
data, except for $\sigma$, the torus polar height parameter, which is
large in both cases. The median values of $\sigma$ are around 60
degrees (measured from the equatorial plane).

In the IR-only fit, several posteriors are bimodal (e.g. $\tau_V$, the
optical depth of a single dust clouds at visual, or $q$, the index of
the power-law $1/r^q$ that describes the radial distribution of clouds
in the torus). It should be remarked that the IR-only fit results in an SED with a significantly stronger flux than the typical SED of ELIRGs \citep{Tsai:2015qf} between 15 and 60$\mu$m. We must point out that the fit to IR data alone runs the risk of over-fitting, since the combined torus+blackbody model has seven free parameters, while there are only five data points fitted. In fact, we artificially set the number of degrees of freedom to unity in this case, to avoid division by a negative number when computing the reduced $\chi^2$. In the all-data fit, with 10 data points, this risk is eliminated.

In the all-data fit, the temperature of the BB component appears
narrowly constrained at 119.5$_{-49.5}^{+14.5}$~K. This value however
is higher than in most other Hot DOGs. \cite{Fan:2016zr} decomposed the
SEDs of 20 Hot DOGs with available WISE data, and found for $T_{\rm\,BB}$ a
range of median values 45--95~K, with their median being 72~K. Preliminary analysis of the 130
sources in the sample of Tsai et al. (in prep.) have good to acceptable
fits, and their BB temperatures span median values 20--126~K, with
their median being 69~K. We believe that the comparatively high BB
temperature in our case arises as an artifact of the lack of data
between 11\mic\ and 160\mic, i.e. our SED currently has no data points
that could help constrain the position of the BB peak. Future studies of the SED of this interesting
source with additional data points between 11 and
160\mic\ will allow to improve the constraints on the parameters.

\begin{table}
\begin{center}
\caption[]{Results of the SED fitting.}
  \begin{tabular}{lrrrr}
    \hline
    \hline
    Parameter & \multicolumn{2}{c}{MAP value} & \multicolumn{2}{c}{Posterior median$\pm 1\sigma$}\\
              & all data & IR data & all data & IR data\\
    \hline\\[-10pt]
    \noalign{\smallskip}
    $i$ (degrees)      &  64.1 &  64.5 &  69.9$_{-15.9}^{+13.3}$ & 66.8$_{-17.3}^{+16.4}$ \\[3pt]
    $\tau_{\rm\,V}$           &  11.7 &  34.1 &  12.6$_{-1.8}^{+2.4}$   & 27.8$_{-12.5}^{+7.3}$  \\[3pt]
    $q$                &   3.0 &   0.4 &   2.6$_{-0.6}^{+0.4}$   &  1.3$_{-0.8}^{+1.5}$   \\[3pt]
    $N_0$              &  14.2 &   6.8 &  13.5$_{-1.3}^{+1.1}$   & 10.6$_{-4.0}^{+3.0}$   \\[3pt]
    $\sigma$ (degrees) &  67.9 &  53.9 &  65.5$_{-5.3}^{+3.3}$   & 60.4$_{-12.4}^{+8.2}$  \\[3pt]
    $Y$                &   8.5 &  51.0 &   6.7$_{-1.7}^{+11.0}$  & 61.6$_{-31.3}^{+28.9}$ \\[3pt]
    $T_{\rm\,BB}$ (K)       & 125.3 & 125.3 & 119.5$_{-49.5}^{+14.5}$ & 82.0$_{-36.0}^{+44.0}$ \\
 \noalign{\smallskip}
   \hline
  \end{tabular}
  \tablecomments{MAP values, posterior medians, and $\pm1\sigma$ confidence intervals around them,
    of all model parameters, for the all-data and IR-data only models shown in Fig.~\ref{fig:fitting}.}
  \label{tab:fitting}
\end{center}
\end{table}

\begin{figure*}
  \begin{center}
\hspace{-5.5cm}
\includegraphics[height=5cm,angle=90]{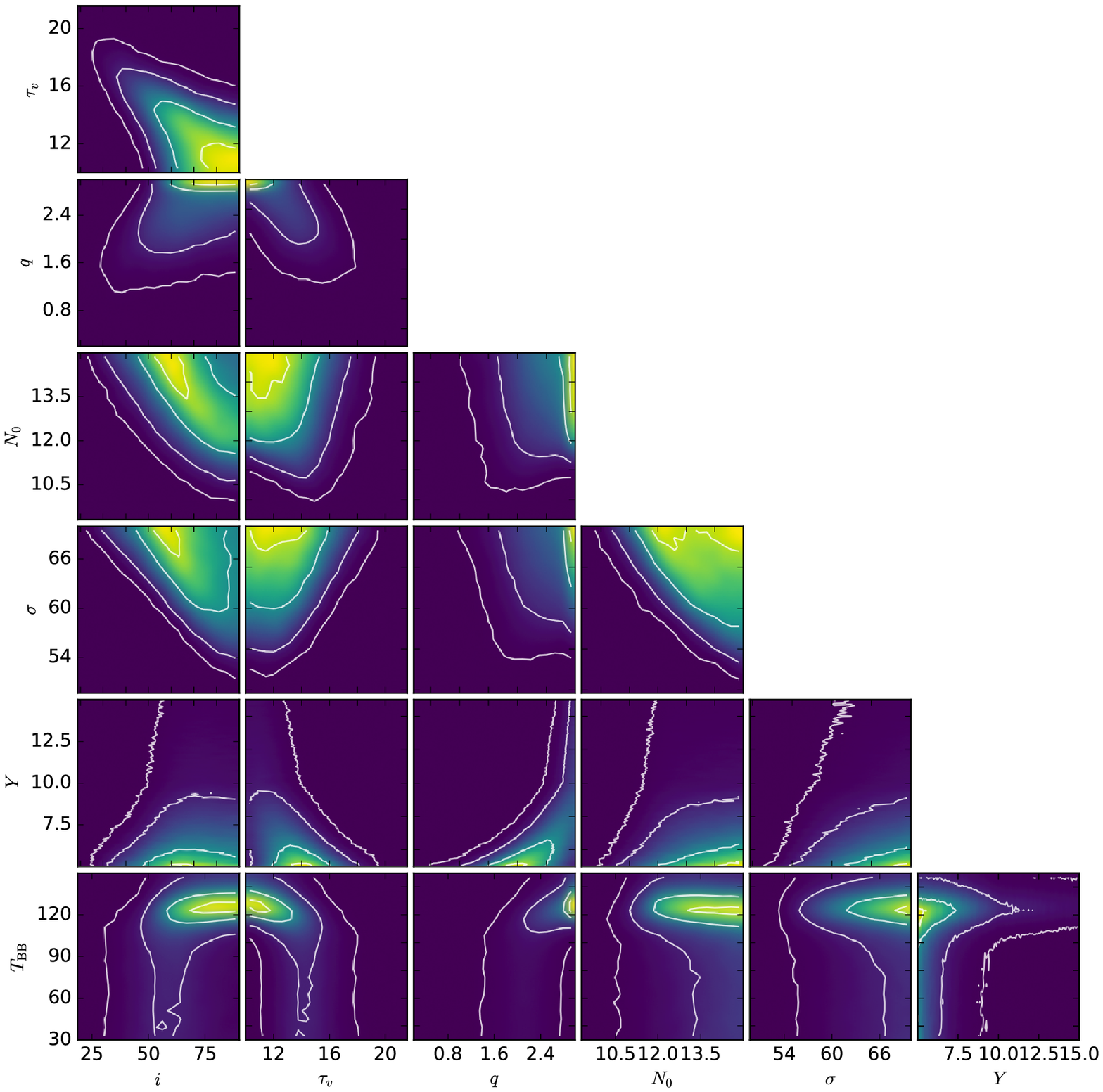}
    \caption{Confidence plot for the SED fit carried out using the complete data set. The white contours are at 1, 10, 50, 90 percent of the peak pixel in each panel.}
    \label{fig:contours_clumpy}
  \end{center}
\end{figure*}

\section{The intrinsic X-ray luminosity of WISE\,1814+3412, WISE\,2207+1939 and WISE\,2357+0328}\label{sect:appendix2}
WISE\,1814+3412, WISE\,2207+1939 and WISE\,2357+0328 were observed by {\it NuSTAR} and {\it XMM-Newton} \citep{Stern:2014kx}, but were either weak or completely undetected by {\it XMM-Newton}. In order to correct the observed X-ray luminosity for absorption we used the following approach.
We adopted as an X-ray spectral template the X-ray spectrum of WISE\,J1036+0449, considering the \textsc{MYTorus} model discussed in $\S$\ref{sect:specAnalysis}. We fixed the value of the line-of-sight column density to that obtained from $E(B-V)$ using Eq.\,\ref{eq:NHebv} (see Table\,\ref{tab:NH_Lx_hotdogs}), set the value of the redshift to that reported in \cite{Stern:2014kx}, and calculated the correction factor in the 2--10\,keV band by changing the inclination angle to $30^{\circ}$. We then extrapolated, by using the X-ray spectral template, the expected luminosity in the rest frame 2--10\,keV band. The values of the luminosities we obtained for these three objects are reported in Table\,\ref{tab:NH_Lx_hotdogs}.

\end{document}